\documentclass[10pt,prl,amsfonts,amssymb,amsmath,aps,notitlepage,twocolumn]{revtex4-2}
\usepackage{graphicx}
\usepackage{subfigure}
\usepackage{amsmath}
\usepackage{color}
\usepackage{hyperref}

\def\ket#1{\lvert\nobreak#1\nobreak\rangle}
\def\bra#1{\langle\nobreak#1\nobreak\rvert}

\def\adag{\hat{a}^{\dag}{}}
\def\a{\hat{a}}
\def\normord#1{\mathop{:}\nolimits\!#1\!\mathop{:}\nolimits}

\newcommand{\op}[1]{\hat{#1}}

\newcommand{\abs}[1]{\lvert #1 \rvert}

\begin{document}

\title{Defining the semiclassical limit of the quantum Rabi Hamiltonian}
\author{E. K. Twyeffort Irish}
\affiliation{School of Physics and Astronomy, University of Southampton, Highfield, Southampton, SO17 1BJ, United Kingdom}
\email{e.k.twyeffort@soton.ac.uk}
\author{A. D. Armour}
\affiliation{School of Physics and Astronomy and Centre for the Mathematics and Theoretical Physics of Quantum Non-Equilibrium Systems, University of Nottingham, Nottingham NG7 2RD, United Kingdom}
\date{\today}

\begin{abstract}
The crossover from quantum to semiclassical behavior in the seminal Rabi model of light-matter interaction still, surprisingly, lacks a complete and rigorous understanding. A formalism for deriving the semiclassical model directly from the quantum Hamiltonian is developed here. Working in a displaced Fock-state basis $\ket{\alpha, n}$, the semiclassical limit is obtained by taking $\lvert \alpha \rvert \to \infty$ and the coupling to zero. This resolves the discrepancy between coherent-state dynamics and semiclassical Rabi oscillations in both standard and ultrastrong coupling/driving regimes. Furthermore, it provides a  framework for studying the quantum-to-semiclassical transition, with potential applications in quantum technologies.
\end{abstract}
\begin{flushleft}

\end{flushleft}

\maketitle

Studies of the physics of a two-level system interacting with an electromagnetic field, as typified by the semiclassical and quantum Rabi models, date back to the 1960s~\cite{Jaynes1963,Shore1993,Braak2016}. Both models have been studied extensively and their predictions are well known. Likewise, it is well known that taking the limit of large photon numbers in the quantum model does not straightforwardly reproduce the semiclassical results, in an apparent contradiction of the correspondence principle~\cite{GerryKnight,MeystreQuantumOptics,BermanMalinovsky,MilonniIntro,ScullyZubairy}. Despite its long and illustrious history, this puzzle in quantum optics has yet to be resolved in a way that is mathematically and physically satisfactory.

For a field alone, defining the transition from quantum to classical is straightforward. The coherent state is known to be the `most classical' of the quantum field states, and its behavior becomes more classical as the average number of photons increases~\cite{Glauber1963}. When the field is coupled to a discrete quantum system, however, the question of correspondence between the quantum and classical models for the field becomes more complicated~\footnote{While it is also possible to consider the question of a classical limit for both components of the system, we focus here on the quantum-to-classical transition of the field alone whilst retaining the quantum nature of the discrete system.}. The semiclassical Rabi model within the rotating-wave approximation (RWA) predicts simple sinusoidal Rabi oscillations of the two-level system. In the corresponding quantum model, taking the field to be in a `classical' coherent state famously produces complex collapse and revival dynamics~\cite{Eberly1980,Narozhny1981}. It is instead the highly nonclassical photon number states, known as Fock states, that lead to sinusoidal oscillations resembling the predictions of the semiclassical theory~\cite{MilonniIntro,GrynbergIntro,GerryKnight,GarrisonChiao,MeystreQuantumOptics}.

Further discrepancies between the quantum and semiclassical results appear in parameter regimes beyond the validity of the RWA, particularly when the quantum coupling and the classical drive amplitude become large. For a high-frequency field, the semiclassical Hamiltonian may be written in terms of Bessel functions that depend on the drive strength~\cite{Shirley1965,Pegg1973, Ashhab2007, Lu2012}, while the quantum energy levels are characterized by Laguerre polynomials in the coupling strength~\cite{Irish2005, Irish2007, Ashhab2007}. An asymptotic relationship between the Laugerre polynomials and the Bessel functions is often invoked to reconcile the quantum and semiclassical predictions~\cite{Plata1993,Neu1996,Grifoni1998}; however, as discussed later, this is questionable on both mathematical and physical grounds. A more rigorous approach requires the assumption of certain statistical properties for the quantum field, resembling those of a coherent state, in order to reproduce the semiclassical results~\cite{Shirley1965, Polonsky1965, AtomPhotonInteractions}. Comparing this with the RWA regime, where coherent states lead to highly non-classical dynamics, highlights a further discrepancy in the existing understanding of the quantum-to-semiclassical correspondence. 
 
With the rise of quantum technology, the distinction between quantum and classical behavior of a field interacting with a two-level system, or qubit, is freighted with practical significance. Engineered quantum devices now routinely operate in regimes where strong single-photon coupling at the quantum level is readily achievable~\cite{Hofheinz2008,Kurizki2015,Blais2021}. Both ultrastrong classical driving~\cite{Nakamura2001,Oliver2005,Wilson2007,Pirkkalainen2013} and ultrastrong quantum coupling~\cite{Niemczyk2010,FornDiaz2010,FornDiaz2019} have been experimentally demonstrated. These achievements open up the possibility of studying the quantum-to-semiclassical transition in unprecedented detail.

In this Letter, we develop a methodology that resolves the question of how to reconcile the quantum and semiclassical predictions. Applying a unitary transformation often used in the ultrastrong coupling regime, we show that the quantum Rabi Hamiltonian may be recast in terms of operator-valued Bessel functions. The appearance of normal ordering in this expression suggests a connection to the semiclassical limit by way of coherent states. Rather than working with coherent states alone, we write the Hamiltonian in terms of displaced Fock states, an orthonormal basis set that serves as a generalization of the coherent states. Taking the displacement amplitude to infinity and the coupling strength to zero, keeping their product finite, recovers the semiclassical Rabi Hamiltonian while preserving the full quantum Hilbert space structure. The same technique may be applied equally well to the untransformed Rabi model, with or without the RWA; however, working in the transformed basis exposes the central importance of the small-coupling limit in the quantum-to-semiclassical transition. We argue that this constitutes a general formalism for defining the semiclassical limit at the Hamiltonian level that is unambiguous, mathematically rigorous, and physically intuitive. What is more, it provides a mathematical framework that will allow the transition to be studied in detail.

The semiclassical Rabi Hamiltonian is
\begin{equation}
\op{H}_{sc}(t) = \tfrac{1}{2} \Omega \op{\sigma}_z + 2A \op{\sigma}_x \cos \omega_0 t ,
\label{Hsc}
\end{equation}
where $\Omega$ is the two-level system frequency, $\op{\sigma}_{x,z}$ are Pauli matrices describing the two-level system, $\omega_0$ is the field frequency, and $A$ is the classical drive amplitude~\footnote{The notation here follows the standard conventions of quantum optics. In circuit QED, the bare energy of the two-level system is frequently defined in the $\op{\sigma}_x$ basis; this differs from the usual quantum optics form by a rotation about $\op{\sigma}_y$.}. 

In the case of strong driving at high frequency, the dynamics of the two-level system exhibits a Bessel-function dependence on the drive amplitude. This result may be obtained by several means, of which Shirley's application of Floquet theory~\cite{Shirley1965} is perhaps the best known. For our purposes, however, the most useful approach is a unitary transformation technique~\cite{Pegg1973, Ashhab2007, Lu2012}. The derivation is briefly outlined here; a full version may be found in the Supplemental Material~\cite{[{See Supplemental Material at }][{ for derivations and a technical discussion of the limiting procedure in the Fock-state basis}]supp}. A transformation is made to a rotating frame with the operator 
\begin{equation}
\op{U}_{sc}(t) = \exp[-i(2A/\omega_0) \op{\sigma}_x \sin \omega_0 t] ,
\label{Usc}
\end{equation} 
which represents the exact solution for the time-evolution operator with $\Omega = 0$~\cite{Castanos2019}. Expanding in terms of Bessel functions, the Hamiltonian becomes
\begin{equation}
\begin{split}
&\tilde{H}_{sc}(t) = \tfrac{1}{2} \Omega \op{\sigma}_z J_0(4A/\omega_0) \\
& \quad + \tfrac{1}{2} \Omega \sum_{p=1}^{\infty} \op{\sigma}_z (-\op{\sigma}_x)^p J_{p}(4A/\omega_0)[e^{ip\omega_0 t} + (-1)^p e^{-ip\omega_0 t}].
\end{split}
\label{scBessel}
\end{equation}
Various approximation schemes may then be employed to derive solutions of the transformed Hamiltonian~\cite{Pegg1973,Ashhab2007,Hausinger2010b,Lu2012}. To lowest order (i.e., neglecting the time-dependent terms~\cite{Pegg1973,Ashhab2007}), the frequency $\Omega$ of the two-level system is renormalized by the coupling to the field, becoming $\Omega_r^{sc} = \Omega J_0(4A/\omega_0)$. 

An analogous approach can be used to study the quantum version of the Rabi Hamiltonian,
\begin{equation}
\op{H}_q = \omega_0 \adag \a + \tfrac{1}{2} \Omega \op{\sigma}_z + \lambda \op{\sigma}_x (\adag + \a) ,
\label{Hrabi}
\end{equation}
where $\adag (\a)$ is the raising (lowering) operator for the quantum field and $\lambda$ is the coupling strength between the two-level system and the field~\footnote{The validity of the Rabi model as an approximation to the underlying light-matter interaction, on the grounds of gauge invariance, has been the subject of recent debate. This question is, however, beyond the scope of the present work; we simply take the quantum Rabi Hamiltonian as a given.}. Solving the $\Omega = 0$ case yields the spin-dependent displacement transformation~\cite{Irish2005, Irish2007}
\begin{equation}
\op{D}\left(-\tfrac{\lambda}{\omega_0} \op{\sigma}_x \right) = \exp \left[-\tfrac{\lambda}{\omega_0} \op{\sigma}_x (\adag - \a) \right] .
\label{disp}
\end{equation}
Under this transformation, the matrix elements of the quantum Hamiltonian in the Fock-state basis become~\cite{Irish2005, Irish2007} 
\begin{equation}
\begin{split}
\langle n+k &\vert \op{D}^{\dag}\op{H}_q \op{D}\vert n \rangle = \left(n \omega_0 - \tfrac{\lambda^2}{\omega_0} \right) \delta_{k,0} \\
& + \tfrac{1}{2} \Omega e^{-2 \lambda^2/\omega_0^2} \bigl(-\tfrac{2\lambda}{\omega_0}\bigr)^k \sqrt{\tfrac{n!}{(n+k)!}}  L_n^k\bigl(\tfrac{4\lambda^2}{\omega_0^2}\bigr) \op{\sigma}_z \op{\sigma}_x^k ,
\end{split}
\label{BesselHam_numbasis}
\end{equation}
for $k = 0, 1, 2, \dots$ \footnote{The corresponding matrix elements within the interaction picture with respect to the field are given in Eq.~(S.36) of the Supplementary Material~\cite{supp}.}. Again taking a lowest-order approximation~\footnote{For the time-independent Hamiltonian $\op{H}_q$ in the lab frame, this approximation may be obtained as the lowest order in degenerate perturbation theory. Alternatively, in the interaction picture with respect to the field (Eq.~(S.36) of the Supplementary Material~\cite{supp}), the approximation consists of neglecting the time-dependent terms, which is directly equivalent to the lowest-order semiclassical approximation mentioned in the previous paragraph.}, the renormalized frequency of the two-level system $\Omega_r^q = \Omega e^{-2\lambda^2/\omega_0^2} L_n(4 \lambda^2/\omega_0^2)$ now depends on the state $\ket{n}$ of the field. The quantum model is characterized by Laguerre polynomials in place of the Bessel functions of the semiclassical model.
 
According to the broadly accepted interpretation of the correspondence principle, the predictions of the semiclassical and quantum models should agree in the limit of large photon numbers~\cite{*[{}] [{, p. 197}] Shankar, Brune1996}. In the literature, a common approach to reconciling the quantum and semiclassical Rabi predictions is to take $n \to \infty$ and apply the asymptotic relation~\cite{GradsteinRyzhik} $\lim_{n \to \infty}n^{-p}{L_n^{p}(x/n)} = x^{-p/2} J_p(2 \sqrt{x})$~\cite{Plata1993,Neu1996,Grifoni1998}. Provided that $\lambda$ is scaled as $A/\sqrt{n}$, where $A$ is identified as the classical drive amplitude, the renormalized frequencies found above become mathematically equivalent. However, simply comparing the frequencies of the two-level system derived from lowest-order approximations is far from a complete correspondence. Attempting to apply a similar argument to the Hamiltonian itself results in both mathematical and conceptual conundrums, as discussed later.

As we now show, a more transparent and rigorous connection between the quantum and semiclassical equations at the Hamiltonian level can be made. The similarities are emphasised by working in a rotating frame with respect to the field. By putting the displacement operator [Eq.~\eqref{disp}] into normal-ordered form~\cite{Armour2013, AtomPhotonInteractions}, the transformed Rabi Hamiltonian may, after some algebra (see Supplemental Material~\cite{[{See Supplemental Material at }][{ for derivations and a technical discussion of the limiting procedure in the Fock-state basis}]supp} for details), be written as
\begin{equation}
\begin{split}
\tilde{H}_q(t) &= -\tfrac{\lambda^2}{\omega_0} + \tfrac{1}{2} \Omega e^{-2 \lambda^2/\omega_0^2} \op{\sigma}_z \normord{J_0(4\lambda \sqrt{\adag \a}/\omega_0)} \\
& \quad + \tfrac{1}{2} \Omega e^{-2 \lambda^2/\omega_0^2} \op{\sigma}_z \sum_{p=1}^{\infty} (-\op{\sigma}_x)^p  \normord{\frac{J_{p}(4\lambda \sqrt{\adag \a}/\omega_0)}{(\sqrt{\adag \a})^{p}} \\
& \qquad \times[e^{ip \omega_0 t}\adag^{p} + (-1)^p e^{-i p \omega_0 t}\a^{p}]} ,
\end{split}
\label{BesselHam}
\end{equation}
where $\normord{{}}$ denotes normal-ordering without the use of commutators, e.g. $\normord{\a \adag} = \adag \a$. This represents an expansion of the transformed Hamiltonian in terms of multi-photon transitions within the displaced basis, with temporal frequencies determined by the number of photons exchanged. Up to this point, no approximations have been made; Eq.~\eqref{BesselHam} is completely equivalent to the original Rabi Hamiltonian. The form of this equation is, to our knowledge, a new result. Its similarity to the semiclassical Hamiltonian in Eq.~\eqref{scBessel} is immediately evident. 

At this point, the textbook recipe for reducing the quantum Hamiltonian to its semiclassical counterpart dictates replacing the quantum field operators $\a$ and $\adag$ by their classical expectation values $\alpha$ and $\alpha^*$ and identifying the classical drive amplitude $A$ with $\lambda \lvert \alpha \rvert = \lambda \sqrt{\bar{n}}$, where $\bar{n}$ is the average photon number~\cite{Jaynes1963,AllenEberly,KlimovChumakov,Ashhab2017,MilonniIntro}. It is readily seen that the untransformed semiclassical Hamiltonian \eqref{Hsc} may be obtained from its quantum counterpart \eqref{Hrabi} in this way. However, applying this recipe to the transformed Hamiltonian~\eqref{BesselHam} reveals a problem: the corresponding semiclassical model~\eqref{scBessel} is not exactly reproduced. The two differ by a factor of $e^{-2 \lambda^2/\omega_0^2}$ and a constant term $-\lambda^2/\omega_0$, which originate from the noncommutativity of $\a$ and $\adag$ and thus are purely quantum effects that should not persist in the semiclassical limit~\footnote{The appearance of the exponential term further depends on the ordering of the field operators; see the Supplemental Material~\cite{supp} for an alternative derivation with a different operator ordering, which gives rise to a different semiclassical expression.}. 

We propose a more rigorous procedure for taking the semiclassical limit, inspired by an approach introduced by Mollow~\cite{Mollow1975} for calculating radiation scattering and later discussed by Pegg~\cite{Pegg1980} and used by Knight and Radmore~\cite{Knight1982} and Berman and Ooi~\cite{Berman2014} to study coherent-state collapse and revival dynamics in the Jaynes-Cummings model. A unitary transformation $\op{D}(\alpha) = \exp[\alpha \adag - \alpha^* \a)]$ is applied directly to the Hamiltonian. This generates a displacement of the field, which may be interpreted as a classical drive~\cite{Mollow1975}. The vacuum field state $\ket{0}$ in this picture corresponds to the coherent state $\ket{\alpha}$ in the original basis. Mathematically, this is equivalent to writing the Hamiltonian in the displaced Fock-state basis $\ket{\alpha, n} \equiv \op{D}(\alpha) \ket{n}$. The quantum Rabi Hamiltonian (in the rotating frame) transforms as
\begin{equation}
\begin{split}
\op{D}^{\dag}(\alpha) \op{H}_q(t) \op{D}(\alpha) &= \tfrac{1}{2} \Omega \op{\sigma}_z + \lambda \op{\sigma}_x(e^{i \omega_0 t}\alpha^* + e^{-i \omega_0 t}\alpha) \\
& \quad + \lambda \op{\sigma}_x(e^{i \omega_0 t}\adag + e^{-i \omega_0 t}\a) .
\end{split}
\end{equation}
The coupling splits into two terms, the first of which is the standard semiclassical driving term while the second is the quantum interaction term. Taking the limit $\lambda \to 0$ while letting $\alpha \to \infty$ eliminates the quantum coupling term, reproducing the semiclassical Hamiltonian~\footnote{This result is implicit in the work of Knight and Radmore~\cite{Knight1982}, although the focus there is on the interplay between the classical and quantum interaction terms in the dynamics of the two-level system. Pegg~\cite{Pegg1980} mentions that neglecting the quantum term gives the semiclassical results, without consideration of the conditions under which such an approximation might be valid.} \footnote{The phase of $\alpha$ determines the phase of the sinusoidal classical drive; for example, choosing $\alpha$ real gives $\cos \omega_0 t$ as in Eq.~\eqref{Hsc}}.

Less trivially, the same idea may be applied to the Bessel function form of the quantum Hamiltonian, Eq.~\eqref{BesselHam}. The matrix elements $\tilde{H}^{n+k,n}_q(t) = {\langle \alpha,~n+k \vert \tilde{H}_q(t) \vert \alpha, n \rangle}$ ($k = 0,1,\dots$) are given by (see derivation in Supplemental Material~\cite{[{See Supplemental Material at }][{ for derivations and a technical discussion of the limiting procedure in the Fock-state basis}]supp}) 
\begin{widetext}
\begin{equation}
\label{BesselHam_cohbasis}
\begin{split}
\tilde{H}^{n+k,n}_q(t) &= -\tfrac{\lambda^2}{\omega_0} \delta_{k,0} + \tfrac{1}{2} \Omega e^{-2 \lambda^2/\omega_0^2} \bigl(-\tfrac{2\lambda}{\omega_0}\bigr)^k \sqrt{\tfrac{n!}{(n+k)!}}  L_n^k\bigl(\tfrac{4\lambda^2}{\omega_0^2}\bigr) \times \biggl\{ \op{\sigma}_z \Bigl(\tfrac{\alpha}{\abs{\alpha}} \Bigr)^{k} J_k(4\lambda \abs{\alpha}/\omega_0) \\
& \qquad + 
 \sum_{p=1}^{\infty} \op{\sigma}_z (-\op{\sigma}_x)^p \biggl[(-1)^k e^{ip \omega_0  t}\Bigl(\tfrac{\alpha^*}{\abs{\alpha}} \Bigr)^{p-k}J_{p-k}\bigl(\tfrac{4\lambda \abs{\alpha}}{\omega_0}\bigr) + (-1)^p e^{-ip \omega_0 t} \Bigl(\tfrac{\alpha}{\abs{\alpha}} \Bigr)^{p+k} J_{p+k}\bigl(\tfrac{4\lambda \abs{\alpha}}{\omega_0}\bigr) \biggr] \biggr\}.
\end{split}
\end{equation}
\end{widetext}
Both quantum and semiclassical features may be identified in this expression. The Laguerre polynomials arise from the quantum model: as $\lvert \alpha \rvert \to 0$, $J_k(4 \lambda \lvert \alpha \rvert / \omega_0) \to \delta_{k,0}$ and the Hamiltonian in the standard Fock-state basis is recovered. The Bessel functions, as previously discussed, are characteristic of semiclassical behavior. 

Taking the limit $\lambda \to 0, \lvert \alpha \rvert \to \infty$ with $\lambda \lvert \alpha \rvert$ held fixed, the off-diagonal terms of Eq.~\eqref{BesselHam_cohbasis} vanish (see Supplemental Material~\cite{[{See Supplemental Material at }][{ for derivations and a technical discussion of the limiting procedure in the Fock-state basis}]supp}) and Eq.~\eqref{BesselHam_cohbasis} reduces to a tensor product of the semiclassical Hamiltonian with the identity operator $\op{I}_{f}$ for the quantum field:
\begin{equation}
\label{sclimit}
\tilde{H}_q(t) \to \tilde{H}_{sc}(t) \otimes \sum_{n=0}^{\infty} \ket{\alpha,n}\bra{\alpha,n} = \tilde{H}_{sc}(t) \otimes \op{I}_{f} .
\end{equation}
The full Hilbert space structure of the quantum model is preserved, but the two-level system now obeys an effective semiclassical Hamiltonian independent of the quantum state of the field. 

Based on these results, we propose a new recipe for reducing the quantum Rabi Hamiltonian to the corresponding semiclassical model:
\begin{enumerate}
\item Transform to a rotating frame with respect to the field mode.
\item Expand in the displaced Fock-state basis $\ket{\alpha, n}$.
\item Take the limit $\lambda \to 0, \lvert \alpha \rvert \to \infty$ such that $\lambda \lvert \alpha \rvert$ remains constant. The semiclassical drive amplitude $A$ corresponds to $\lambda \lvert \alpha \rvert$.
\end{enumerate}
The semiclassical Hamiltonian is thus obtained directly from the quantum Hamiltonian, without specifying a particular initial state for the quantum field or imposing assumptions about its statistical properties. Since the Hilbert space structure is maintained and the procedure involves a well-defined mathematical limit, this approach opens up the possibility of studying the crossover from quantum to semiclassical behavior by examining quantum perturbations to the semiclassical Hamiltonian. 

The two key ingredients are the choice of basis states and the form of the mathematical limits. As coherent states are the most classical states of a quantized field, it is natural to expect them to be involved. The displaced Fock states $\ket{\alpha,n}$ may be viewed as interpolating between the semiclassical coherent states $\ket{\alpha}$ and the eigenstates $\ket{n}$ of the quantized field. Unlike the overcomplete set of coherent states, the set $\{\ket{\alpha, n}\}$ with fixed $\alpha$ forms an orthonormal basis set with properties similar to the Fock state basis. For $n>0$ these states are distinctly nonclassical. Nevertheless, we argue that the displaced Fock states constitute the correct basis for carrying out the semiclassical limit~\footnote{See Supplemental Material~\cite{[{See Supplemental Material at }][{ for derivations and a technical discussion of the limiting procedure in the Fock-state basis}]supp} for a further, more technical discussion of this point, based on the structure of the Hamiltonian.}. 

To support this claim, let us first examine the time evolution of $\ket{\alpha, n}$. Upon transforming back out of the rotating frame with respect to the field Hamiltonian, the states become time dependent: $e^{-i \omega_0 t \adag \a} \ket{\alpha, n} = e^{-i n \omega_0 t} \ket{\alpha e^{-i \omega_0 t}, n}$. Defining a general time-dependent state vector for the field $\ket{\psi(t)} = \sum_m c_m(t) e^{-i m \omega_0 t} \ket{\alpha e^{-i \omega_0 t}, m}$ and a spin state $\ket{\phi}$, the time evolution generated by the correspondingly transformed Hamiltonian may be expressed in terms of the matrix elements given in Eq.~\eqref{BesselHam_cohbasis}:
\begin{equation}
\begin{split}
&\sum_{m=0}^{\infty}  i \dot{c}_m(t) \ket{\alpha e^{-i \omega_0 t}, m} \ket{\phi} \\
& ~= \sum_{m=0}^{\infty} c_m(t) \tilde{H}^{m,m}_q(t) \ket{\alpha e^{-i \omega_0 t}, m} \ket{\phi}\\
& \quad + \sum_{m=0}^{\infty} \sum_{k=1}^{\infty} c_m(t) e^{-i k \omega_0 t} \tilde{H}^{m+k,m}_q(t) \ket{\alpha e^{-i \omega_0 t}, m+k} \ket{\phi}\\
& \quad + \sum_{m=0}^{\infty} \sum_{k=1}^{m} c_m(t) e^{i k \omega_0 t} \tilde{H}^{m-k,m}_q(t) \ket{\alpha e^{-i \omega_0 t}, m-k}\ket{\phi}.
\end{split}
\end{equation}
An initial state $\ket{\alpha e^{-i \omega_0 t}, n}$ will evolve over time into a superposition of displaced Fock states. In the semiclassical limit, however, the off-diagonal terms of $\tilde{H}_q$ vanish and the basis states $\ket{\alpha e^{-i \omega_0 t}, m}$ become uncoupled. A field state $\ket{\alpha e^{-i \omega_0 t}, n}$ then undergoes intrinsic time evolution corresponding to a rotation in phase space -- such that the expectation values of operators obey the classical harmonic oscillator equations of motion -- but its amplitude remains constant. Meanwhile, the spin obeys an effective Hamiltonian that is independent of the quantum state of the field. The spin and field remain in a separable state at all times: precisely the expected semiclassical behavior.

This does not, however, imply that all of the displaced Fock states may be considered equally `classical'. The dispersion of the position and momentum operators in $\ket{\alpha, n}$ scales as $n$, so states with $n>0$ exhibit greater quantum fluctuations than the minimum imposed by the uncertainty principle; they may also have negative-valued Wigner functions, another hallmark of nonclassical behavior. Off-diagonal terms in $\tilde{H}_q$ that couple $\ket{\alpha, n}$ with $\ket{\alpha, n+k}$ scale as $(\lambda \sqrt{n})^k$. This suggests that, for finite values of $\alpha$ and $\lambda$, leakage into different states happens on faster timescales for larger $n$. Hence the semiclassical evolution of the displaced Fock states becomes increasingly fragile against quantum corrections as $n$ increases. 

Turning next to the limits, taking the field amplitude (as measured by the average photon number or coherent-state amplitude) to infinity is widely assumed to correspond to the correct semiclassical limit~\cite{Brune1996, MilonniIntro, GrynbergIntro, BermanMalinovsky, GarrisonChiao, MeystreQuantumOptics, Everitt2009}. In our formalism, this is accounted for by the limit $\lvert \alpha \rvert \to \infty$. (Note, however, that $\alpha$ here serves as a continuous $c$-number variable that parameterizes a unitary transformation and cannot, in general, be identified with the average photon number~\footnote{See Supplementary Material~\cite{[{See Supplemental Material at }][{ for derivations and a technical discussion of the limiting procedure in the Fock-state basis}]supp} for further discussion of this distinction and its importance.}.) By contrast, the small-coupling limit $\lambda \to 0$ is widely neglected in the literature, apart from an occasional mention that this limit allows the coherent-state dynamics in the Jaynes-Cummings model to be reconciled with the semiclassical predictions (e.g.~\cite{HarocheRaimond, BermanMalinovsky,Hohenester}; a more careful discussion is found in~\cite{Larson2021}). Working in the transformed basis defined by Eq.~\eqref{disp} reveals the central necessity of this limit. It is, in fact, the $\lambda \to 0$ limit that eliminates the quantum terms from the Hamiltonian; taking $\lvert \alpha \rvert \to \infty$ is only needed to ensure that the classical drive amplitude does not vanish. The physical interpretation is clear and intuitive: in the semiclassical limit, not only must the field become classical, but the interaction of the two-level system with individual photons must become negligible. 

\begin{figure}[tb]
		\includegraphics[width=8.6cm]{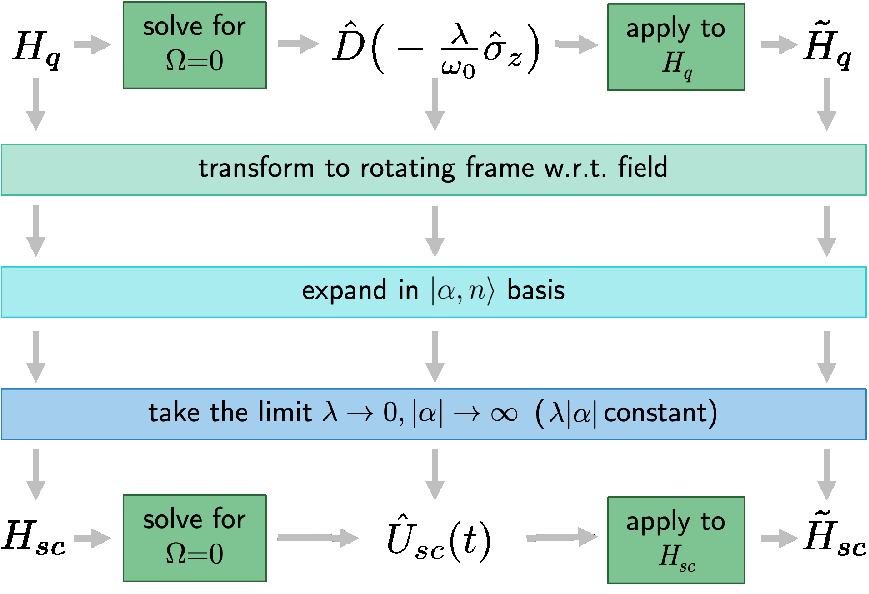}
	\caption{Schematic of the transformations and limits employed here, illustrating the parallels between the quantum and semiclassical cases. Moving from left to right, solving either the quantum or the semiclassical Rabi Hamiltonian with $\Omega = 0$ produces a unitary transformation, which can then be applied to the full Hamiltonian to give $\tilde{H}_q$ or $\tilde{H}_{sc}$, respectively. The procedure for taking the semiclassical limit is shown from top to bottom. This may be applied to the standard Rabi Hamiltonian $H_q$, the transformed version $\tilde{H}_q$, or even the transformation operator itself. In each case, the corresponding semiclassical result is obtained.}
	\label{fig:transformations}
\end{figure}

Considering these limits clarifies the relationship between ultrastrong coupling in the quantum model and ultrastrong driving in the semiclassical model. As the classical drive amplitude is $A = \lambda \lvert \alpha \rvert$, strong driving may be obtained by taking either the quantum coupling $\lambda$ or the field amplitude $\alpha$ (or both) to be large~\cite{Ashhab2017}. Within the theoretical framework established here, $\lambda$ must go to zero in the semiclassical limit. Consequently, a semiclassical limit for ultrastrong quantum coupling cannot, in principle, exist. While the case of strong semiclassical driving parallels that of strong quantum coupling in the sense illustrated in Fig.~\ref{fig:transformations}, the semiclassical drive must be provided by a large amplitude field with a vanishingly small single-photon coupling.  

Intriguingly, the same recipe may be used to derive the semiclassical transformation operator $\op{U}_{sc}(t)$ from the quantum operator $\op{D}[(-\lambda/\omega_0)\op{\sigma}_x]$ (see Supplemental Material~\cite{[{See Supplemental Material at }][{ for derivations and a technical discussion of the limiting procedure in the Fock-state basis}]supp}). This completes the correspondence between the quantum and semiclassical cases, as summarized in Fig.~\ref{fig:transformations}. It furthermore suggests that the procedure developed here for the specific case of the Rabi model may have wider applicability to related models of light-matter interaction~\cite{Larson2021}.

To conclude, we have developed a mathematically rigorous and physically intuitive method for obtaining the semiclassical Rabi model from the underlying quantum model at the Hamiltonian level. The arguments presented here indicate that the semiclassical limit emerges most naturally when the quantum field is expressed in the basis of displaced Fock states. The time evolution of these states converges to the expected semiclassical dynamics when the appropriate mathematical limit is carried out. This approach appears almost trivial when applied to the standard form of the Rabi Hamiltonian. A more compelling case, however, emerges from a transformed model in which the quantum Rabi Hamiltonian is expressed in terms of operator-valued Bessel functions. The derivation of this form, which constitutes a notable result in its own right, parallels the Bessel-function expansion that has long been known for the semiclassical model. 

The formalism presented here resolves the long-standing question in quantum optics theory regarding the emergence of the semiclassical limit from the quantum Rabi model. Importantly, it is equally applicable in both the standard parameter regime (including the Jaynes-Cummings model) and the ultrastrong coupling/driving regimes that have attracted increasing theoretical and experimental interest in recent years. As the full quantum Hilbert space structure is preserved in the process of taking the semiclassical limit, the method provides a natural framework for calculating quantum corrections to the semiclassical dynamics. This will enable studies of the effect of field quantisation on operations where a classical driving field is usually assumed, a situation of considerable experimental relevance in cavity and circuit QED and related quantum technologies.

\bibliography{Rabi_Bessel_paper}

\begin{thebibliography}{62}%
\makeatletter
\providecommand \@ifxundefined [1]{%
 \@ifx{#1\undefined}
}%
\providecommand \@ifnum [1]{%
 \ifnum #1\expandafter \@firstoftwo
 \else \expandafter \@secondoftwo
 \fi
}%
\providecommand \@ifx [1]{%
 \ifx #1\expandafter \@firstoftwo
 \else \expandafter \@secondoftwo
 \fi
}%
\providecommand \natexlab [1]{#1}%
\providecommand \enquote  [1]{``#1''}%
\providecommand \bibnamefont  [1]{#1}%
\providecommand \bibfnamefont [1]{#1}%
\providecommand \citenamefont [1]{#1}%
\providecommand \href@noop [0]{\@secondoftwo}%
\providecommand \href [0]{\begingroup \@sanitize@url \@href}%
\providecommand \@href[1]{\@@startlink{#1}\@@href}%
\providecommand \@@href[1]{\endgroup#1\@@endlink}%
\providecommand \@sanitize@url [0]{\catcode `\\12\catcode `\$12\catcode
  `\&12\catcode `\#12\catcode `\^12\catcode `\_12\catcode `\%12\relax}%
\providecommand \@@startlink[1]{}%
\providecommand \@@endlink[0]{}%
\providecommand \url  [0]{\begingroup\@sanitize@url \@url }%
\providecommand \@url [1]{\endgroup\@href {#1}{\urlprefix }}%
\providecommand \urlprefix  [0]{URL }%
\providecommand \Eprint [0]{\href }%
\providecommand \doibase [0]{https://doi.org/}%
\providecommand \selectlanguage [0]{\@gobble}%
\providecommand \bibinfo  [0]{\@secondoftwo}%
\providecommand \bibfield  [0]{\@secondoftwo}%
\providecommand \translation [1]{[#1]}%
\providecommand \BibitemOpen [0]{}%
\providecommand \bibitemStop [0]{}%
\providecommand \bibitemNoStop [0]{.\EOS\space}%
\providecommand \EOS [0]{\spacefactor3000\relax}%
\providecommand \BibitemShut  [1]{\csname bibitem#1\endcsname}%
\let\auto@bib@innerbib\@empty
\bibitem [{\citenamefont {Jaynes}\ and\ \citenamefont
  {Cummings}(1963)}]{Jaynes1963}%
  \BibitemOpen
  \bibfield  {author} {\bibinfo {author} {\bibfnamefont {E.~T.}\ \bibnamefont
  {Jaynes}}\ and\ \bibinfo {author} {\bibfnamefont {F.~W.}\ \bibnamefont
  {Cummings}},\ }\bibfield  {title} {\bibinfo {title} {Comparison of quantum
  and semiclassical radiation theories with application to the beam maser},\
  }\href {https://doi.org/10.1109/proc.1963.1664} {\bibfield  {journal}
  {\bibinfo  {journal} {Proc. IEEE}\ }\textbf {\bibinfo {volume} {51}},\
  \bibinfo {pages} {89} (\bibinfo {year} {1963})}\BibitemShut {NoStop}%
\bibitem [{\citenamefont {Shore}\ and\ \citenamefont
  {Knight}(1993)}]{Shore1993}%
  \BibitemOpen
  \bibfield  {author} {\bibinfo {author} {\bibfnamefont {B.~W.}\ \bibnamefont
  {Shore}}\ and\ \bibinfo {author} {\bibfnamefont {P.~L.}\ \bibnamefont
  {Knight}},\ }\bibfield  {title} {\bibinfo {title} {The {J}aynes-{C}ummings
  model},\ }\href {https://doi.org/10.1080/09500349314551321} {\bibfield
  {journal} {\bibinfo  {journal} {J. Mod. Opt.}\ }\textbf {\bibinfo {volume}
  {40}},\ \bibinfo {pages} {1195} (\bibinfo {year} {1993})}\BibitemShut
  {NoStop}%
\bibitem [{\citenamefont {Braak}\ \emph {et~al.}(2016)\citenamefont {Braak},
  \citenamefont {Chen}, \citenamefont {Batchelor},\ and\ \citenamefont
  {Solano}}]{Braak2016}%
  \BibitemOpen
  \bibfield  {author} {\bibinfo {author} {\bibfnamefont {D.}~\bibnamefont
  {Braak}}, \bibinfo {author} {\bibfnamefont {Q.~H.}\ \bibnamefont {Chen}},
  \bibinfo {author} {\bibfnamefont {M.~T.}\ \bibnamefont {Batchelor}},\ and\
  \bibinfo {author} {\bibfnamefont {E.}~\bibnamefont {Solano}},\ }\bibfield
  {title} {\bibinfo {title} {Semi-classical and quantum {R}abi models: in
  celebration of 80 years},\ }\href
  {https://doi.org/10.1088/1751-8113/49/30/300301} {\bibfield  {journal}
  {\bibinfo  {journal} {J. Phys. A}\ }\textbf {\bibinfo {volume} {49}},\
  \bibinfo {pages} {300301} (\bibinfo {year} {2016})}\BibitemShut {NoStop}%
\bibitem [{\citenamefont {Gerry}\ and\ \citenamefont
  {Knight}(2004)}]{GerryKnight}%
  \BibitemOpen
  \bibfield  {author} {\bibinfo {author} {\bibfnamefont {C.}~\bibnamefont
  {Gerry}}\ and\ \bibinfo {author} {\bibfnamefont {P.}~\bibnamefont {Knight}},\
  }\href {https://doi.org/10.1017/CBO9780511791239} {\emph {\bibinfo {title}
  {Introductory Quantum Optics}}}\ (\bibinfo  {publisher} {Cambridge University
  Press},\ \bibinfo {address} {Cambridge},\ \bibinfo {year} {2004})\BibitemShut
  {NoStop}%
\bibitem [{\citenamefont {Meystre}(2021)}]{MeystreQuantumOptics}%
  \BibitemOpen
  \bibfield  {author} {\bibinfo {author} {\bibfnamefont {P.}~\bibnamefont
  {Meystre}},\ }\href@noop {} {\emph {\bibinfo {title} {Quantum Optics: Taming
  the Quantum}}}\ (\bibinfo  {publisher} {Springer},\ \bibinfo {address} {Cham,
  Switzerland},\ \bibinfo {year} {2021})\BibitemShut {NoStop}%
\bibitem [{\citenamefont {Berman}\ and\ \citenamefont
  {Malinovsky}(2011)}]{BermanMalinovsky}%
  \BibitemOpen
  \bibfield  {author} {\bibinfo {author} {\bibfnamefont {P.~R.}\ \bibnamefont
  {Berman}}\ and\ \bibinfo {author} {\bibfnamefont {V.~S.}\ \bibnamefont
  {Malinovsky}},\ }\href@noop {} {\emph {\bibinfo {title} {Principles of Laser
  Spectroscopy and Quantum Optics}}}\ (\bibinfo  {publisher} {Princeton
  University Press},\ \bibinfo {address} {Princeton, New Jersey},\ \bibinfo
  {year} {2011})\BibitemShut {NoStop}%
\bibitem [{\citenamefont {Milonni}(2019)}]{MilonniIntro}%
  \BibitemOpen
  \bibfield  {author} {\bibinfo {author} {\bibfnamefont {P.~W.}\ \bibnamefont
  {Milonni}},\ }\href@noop {} {\emph {\bibinfo {title} {An Introduction to
  Quantum Optics and Quantum Fluctuations}}}\ (\bibinfo  {publisher} {Oxford
  University Press},\ \bibinfo {address} {Oxford},\ \bibinfo {year}
  {2019})\BibitemShut {NoStop}%
\bibitem [{\citenamefont {Scully}\ and\ \citenamefont
  {Zubairy}(1997)}]{ScullyZubairy}%
  \BibitemOpen
  \bibfield  {author} {\bibinfo {author} {\bibfnamefont {M.~O.}\ \bibnamefont
  {Scully}}\ and\ \bibinfo {author} {\bibfnamefont {M.~S.}\ \bibnamefont
  {Zubairy}},\ }\href {https://doi.org/10.1017/CBO9780511813993} {\emph
  {\bibinfo {title} {Quantum Optics}}}\ (\bibinfo  {publisher} {Cambridge
  University Press},\ \bibinfo {address} {Cambridge},\ \bibinfo {year}
  {1997})\BibitemShut {NoStop}%
\bibitem [{\citenamefont {Glauber}(1963)}]{Glauber1963}%
  \BibitemOpen
  \bibfield  {author} {\bibinfo {author} {\bibfnamefont {R.~J.}\ \bibnamefont
  {Glauber}},\ }\bibfield  {title} {\bibinfo {title} {Coherent and incoherent
  states of the radiation field},\ }\href
  {https://doi.org/10.1103/PhysRev.131.2766} {\bibfield  {journal} {\bibinfo
  {journal} {Phys. Rev.}\ }\textbf {\bibinfo {volume} {131}},\ \bibinfo {pages}
  {2766} (\bibinfo {year} {1963})}\BibitemShut {NoStop}%
\bibitem [{Note1()}]{Note1}%
  \BibitemOpen
  \bibinfo {note} {While it is also possible to consider the question of a
  classical limit for both components of the system, we focus here on the
  quantum-to-classical transition of the field alone whilst retaining the
  quantum nature of the discrete system.}\BibitemShut {Stop}%
\bibitem [{\citenamefont {Eberly}\ \emph {et~al.}(1980)\citenamefont {Eberly},
  \citenamefont {Narozhny},\ and\ \citenamefont
  {Sanchez-Mondragon}}]{Eberly1980}%
  \BibitemOpen
  \bibfield  {author} {\bibinfo {author} {\bibfnamefont {J.~H.}\ \bibnamefont
  {Eberly}}, \bibinfo {author} {\bibfnamefont {N.~B.}\ \bibnamefont
  {Narozhny}},\ and\ \bibinfo {author} {\bibfnamefont {J.~J.}\ \bibnamefont
  {Sanchez-Mondragon}},\ }\bibfield  {title} {\bibinfo {title} {Periodic
  spontaneous collapse and revival in a simple quantum model},\ }\href
  {https://doi.org/10.1103/PhysRevLett.44.1323} {\bibfield  {journal} {\bibinfo
   {journal} {Phys. Rev. Lett.}\ }\textbf {\bibinfo {volume} {44}},\ \bibinfo
  {pages} {1323} (\bibinfo {year} {1980})}\BibitemShut {NoStop}%
\bibitem [{\citenamefont {Narozhny}\ \emph {et~al.}(1981)\citenamefont
  {Narozhny}, \citenamefont {Sanchez-Mondragon},\ and\ \citenamefont
  {Eberly}}]{Narozhny1981}%
  \BibitemOpen
  \bibfield  {author} {\bibinfo {author} {\bibfnamefont {N.~B.}\ \bibnamefont
  {Narozhny}}, \bibinfo {author} {\bibfnamefont {J.~J.}\ \bibnamefont
  {Sanchez-Mondragon}},\ and\ \bibinfo {author} {\bibfnamefont {J.~H.}\
  \bibnamefont {Eberly}},\ }\bibfield  {title} {\bibinfo {title} {Coherence
  versus incoherence: Collapse and revival in a simple quantum model},\ }\href
  {https://doi.org/10.1103/PhysRevA.23.236} {\bibfield  {journal} {\bibinfo
  {journal} {Phys. Rev. A}\ }\textbf {\bibinfo {volume} {23}},\ \bibinfo
  {pages} {236} (\bibinfo {year} {1981})}\BibitemShut {NoStop}%
\bibitem [{\citenamefont {Grynberg}\ \emph {et~al.}(2010)\citenamefont
  {Grynberg}, \citenamefont {Aspect},\ and\ \citenamefont
  {Fabre}}]{GrynbergIntro}%
  \BibitemOpen
  \bibfield  {author} {\bibinfo {author} {\bibfnamefont {G.}~\bibnamefont
  {Grynberg}}, \bibinfo {author} {\bibfnamefont {A.}~\bibnamefont {Aspect}},\
  and\ \bibinfo {author} {\bibfnamefont {C.}~\bibnamefont {Fabre}},\
  }\href@noop {} {\emph {\bibinfo {title} {Introduction to Quantum Optics: From
  the Semi-Classical Approach to Quantized Light}}}\ (\bibinfo  {publisher}
  {Cambridge University Press},\ \bibinfo {address} {Cambridge},\ \bibinfo
  {year} {2010})\BibitemShut {NoStop}%
\bibitem [{\citenamefont {Garrison}\ and\ \citenamefont
  {Chiao}(2008)}]{GarrisonChiao}%
  \BibitemOpen
  \bibfield  {author} {\bibinfo {author} {\bibfnamefont {J.~C.}\ \bibnamefont
  {Garrison}}\ and\ \bibinfo {author} {\bibfnamefont {R.~Y.}\ \bibnamefont
  {Chiao}},\ }\href@noop {} {\emph {\bibinfo {title} {Quantum Optics}}}\
  (\bibinfo  {publisher} {Oxford University Press},\ \bibinfo {address}
  {Oxford},\ \bibinfo {year} {2008})\BibitemShut {NoStop}%
\bibitem [{\citenamefont {Shirley}(1965)}]{Shirley1965}%
  \BibitemOpen
  \bibfield  {author} {\bibinfo {author} {\bibfnamefont {J.}~\bibnamefont
  {Shirley}},\ }\bibfield  {title} {\bibinfo {title} {Solution of the
  {S}chr\"{o}dinger equation with a {H}amiltonian periodic in time},\ }\href
  {https://doi.org/10.1103/PhysRev.138.B979} {\bibfield  {journal} {\bibinfo
  {journal} {Phys. Rev.}\ }\textbf {\bibinfo {volume} {138}},\ \bibinfo {pages}
  {B979} (\bibinfo {year} {1965})}\BibitemShut {NoStop}%
\bibitem [{\citenamefont {Pegg}\ and\ \citenamefont {Series}(1973)}]{Pegg1973}%
  \BibitemOpen
  \bibfield  {author} {\bibinfo {author} {\bibfnamefont {D.~T.}\ \bibnamefont
  {Pegg}}\ and\ \bibinfo {author} {\bibfnamefont {G.~W.}\ \bibnamefont
  {Series}},\ }\bibfield  {title} {\bibinfo {title} {On the reduction of a
  problem in magnetic resonance},\ }\href
  {https://doi.org/10.1098/RSPA.1973.0026} {\bibfield  {journal} {\bibinfo
  {journal} {Proc. Roy. Soc. A}\ }\textbf {\bibinfo {volume} {332}},\ \bibinfo
  {pages} {281} (\bibinfo {year} {1973})}\BibitemShut {NoStop}%
\bibitem [{\citenamefont {Ashhab}\ \emph {et~al.}(2007)\citenamefont {Ashhab},
  \citenamefont {Johansson}, \citenamefont {Zagoskin},\ and\ \citenamefont
  {Nori}}]{Ashhab2007}%
  \BibitemOpen
  \bibfield  {author} {\bibinfo {author} {\bibfnamefont {S.}~\bibnamefont
  {Ashhab}}, \bibinfo {author} {\bibfnamefont {J.~R.}\ \bibnamefont
  {Johansson}}, \bibinfo {author} {\bibfnamefont {A.~M.}\ \bibnamefont
  {Zagoskin}},\ and\ \bibinfo {author} {\bibfnamefont {F.}~\bibnamefont
  {Nori}},\ }\bibfield  {title} {\bibinfo {title} {Two-level systems driven by
  large-amplitude fields},\ }\href {https://doi.org/10.1103/PhysRevA.75.063414}
  {\bibfield  {journal} {\bibinfo  {journal} {Phys. Rev. A}\ }\textbf {\bibinfo
  {volume} {75}},\ \bibinfo {pages} {063414} (\bibinfo {year}
  {2007})}\BibitemShut {NoStop}%
\bibitem [{\citenamefont {L\"u}\ and\ \citenamefont {Zheng}(2012)}]{Lu2012}%
  \BibitemOpen
  \bibfield  {author} {\bibinfo {author} {\bibfnamefont {Z.}~\bibnamefont
  {L\"u}}\ and\ \bibinfo {author} {\bibfnamefont {H.}~\bibnamefont {Zheng}},\
  }\bibfield  {title} {\bibinfo {title} {Effects of counter-rotating
  interaction on driven tunneling dynamics: Coherent destruction of tunneling
  and {B}loch-{S}iegert shift},\ }\href
  {https://doi.org/10.1103/PhysRevA.86.023831} {\bibfield  {journal} {\bibinfo
  {journal} {Phys. Rev. A}\ }\textbf {\bibinfo {volume} {86}},\ \bibinfo
  {pages} {023831} (\bibinfo {year} {2012})}\BibitemShut {NoStop}%
\bibitem [{\citenamefont {Twyeffort~Irish}\ \emph {et~al.}(2005)\citenamefont
  {Twyeffort~Irish}, \citenamefont {Gea-Banacloche}, \citenamefont {Martin},\
  and\ \citenamefont {Schwab}}]{Irish2005}%
  \BibitemOpen
  \bibfield  {author} {\bibinfo {author} {\bibfnamefont {E.~K.}\ \bibnamefont
  {Twyeffort~Irish}}, \bibinfo {author} {\bibfnamefont {J.}~\bibnamefont
  {Gea-Banacloche}}, \bibinfo {author} {\bibfnamefont {I.}~\bibnamefont
  {Martin}},\ and\ \bibinfo {author} {\bibfnamefont {K.~C.}\ \bibnamefont
  {Schwab}},\ }\bibfield  {title} {\bibinfo {title} {Dynamics of a two-level
  system strongly coupled to a high-frequency quantum oscillator},\ }\href
  {https://doi.org/10.1103/PhysRevB.72.195410} {\bibfield  {journal} {\bibinfo
  {journal} {Phys. Rev. B}\ }\textbf {\bibinfo {volume} {72}},\ \bibinfo
  {pages} {195410} (\bibinfo {year} {2005})}\BibitemShut {NoStop}%
\bibitem [{\citenamefont {Twyeffort~Irish}(2007)}]{Irish2007}%
  \BibitemOpen
  \bibfield  {author} {\bibinfo {author} {\bibfnamefont {E.~K.}\ \bibnamefont
  {Twyeffort~Irish}},\ }\bibfield  {title} {\bibinfo {title} {Generalized
  rotating-wave approximation for arbitrarily large coupling},\ }\href
  {https://doi.org/10.1103/PhysRevLett.99.173601} {\bibfield  {journal}
  {\bibinfo  {journal} {Phys. Rev. Lett.}\ }\textbf {\bibinfo {volume} {99}},\
  \bibinfo {pages} {173601} (\bibinfo {year} {2007})}\BibitemShut {NoStop}%
\bibitem [{\citenamefont {Plata}\ and\ \citenamefont
  {Gomez~Llorente}(1993)}]{Plata1993}%
  \BibitemOpen
  \bibfield  {author} {\bibinfo {author} {\bibfnamefont {J.}~\bibnamefont
  {Plata}}\ and\ \bibinfo {author} {\bibfnamefont {J.~M.}\ \bibnamefont
  {Gomez~Llorente}},\ }\bibfield  {title} {\bibinfo {title} {Control of
  tunneling in an electromagnetic cavity},\ }\href
  {https://doi.org/10.1103/PhysRevA.48.782} {\bibfield  {journal} {\bibinfo
  {journal} {Phys. Rev. A}\ }\textbf {\bibinfo {volume} {48}},\ \bibinfo
  {pages} {782} (\bibinfo {year} {1993})}\BibitemShut {NoStop}%
\bibitem [{\citenamefont {Neu}\ and\ \citenamefont {Silbey}(1996)}]{Neu1996}%
  \BibitemOpen
  \bibfield  {author} {\bibinfo {author} {\bibfnamefont {P.}~\bibnamefont
  {Neu}}\ and\ \bibinfo {author} {\bibfnamefont {R.~J.}\ \bibnamefont
  {Silbey}},\ }\bibfield  {title} {\bibinfo {title} {Tunneling in a cavity},\
  }\href {https://doi.org/10.1103/PhysRevA.54.5323} {\bibfield  {journal}
  {\bibinfo  {journal} {Phys. Rev. A}\ }\textbf {\bibinfo {volume} {54}},\
  \bibinfo {pages} {5323} (\bibinfo {year} {1996})}\BibitemShut {NoStop}%
\bibitem [{\citenamefont {Grifoni}\ and\ \citenamefont
  {H{\"a}nggi}(1998)}]{Grifoni1998}%
  \BibitemOpen
  \bibfield  {author} {\bibinfo {author} {\bibfnamefont {M.}~\bibnamefont
  {Grifoni}}\ and\ \bibinfo {author} {\bibfnamefont {P.}~\bibnamefont
  {H{\"a}nggi}},\ }\bibfield  {title} {\bibinfo {title} {Driven quantum
  tunneling},\ }\href {https://doi.org/10.1016/S0370-1573(98)00022-2}
  {\bibfield  {journal} {\bibinfo  {journal} {Phys. Rep.}\ }\textbf {\bibinfo
  {volume} {304}},\ \bibinfo {pages} {229} (\bibinfo {year}
  {1998})}\BibitemShut {NoStop}%
\bibitem [{\citenamefont {Polonsky}\ and\ \citenamefont
  {Cohen-Tannoudji}(1965)}]{Polonsky1965}%
  \BibitemOpen
  \bibfield  {author} {\bibinfo {author} {\bibfnamefont {N.}~\bibnamefont
  {Polonsky}}\ and\ \bibinfo {author} {\bibfnamefont {C.}~\bibnamefont
  {Cohen-Tannoudji}},\ }\bibfield  {title} {\bibinfo {title}
  {Interpr\'{e}tation quantique de la modulation de fr\'{e}quence},\ }\href
  {https://doi.org/10.1051/jphys:01965002607040900} {\bibfield  {journal}
  {\bibinfo  {journal} {J. Phys. (Paris)}\ }\textbf {\bibinfo {volume} {26}},\
  \bibinfo {pages} {409} (\bibinfo {year} {1965})}\BibitemShut {NoStop}%
\bibitem [{\citenamefont {Cohen-Tannoudji}\ \emph {et~al.}(2004)\citenamefont
  {Cohen-Tannoudji}, \citenamefont {Dupont-Roc},\ and\ \citenamefont
  {Grynberg}}]{AtomPhotonInteractions}%
  \BibitemOpen
  \bibfield  {author} {\bibinfo {author} {\bibfnamefont {C.}~\bibnamefont
  {Cohen-Tannoudji}}, \bibinfo {author} {\bibfnamefont {J.}~\bibnamefont
  {Dupont-Roc}},\ and\ \bibinfo {author} {\bibfnamefont {G.}~\bibnamefont
  {Grynberg}},\ }\href@noop {} {\emph {\bibinfo {title} {Atom-photon
  interactions: basic processes and applications}}}\ (\bibinfo  {publisher}
  {Wiley-VCH},\ \bibinfo {address} {Weinheim},\ \bibinfo {year}
  {2004})\BibitemShut {NoStop}%
\bibitem [{\citenamefont {Hofheinz}\ \emph {et~al.}(2008)\citenamefont
  {Hofheinz}, \citenamefont {Weig}, \citenamefont {Ansmann}, \citenamefont
  {Bialczak}, \citenamefont {Lucero}, \citenamefont {Neeley}, \citenamefont
  {O’Connell}, \citenamefont {Wang}, \citenamefont {Martinis},\ and\
  \citenamefont {Cleland}}]{Hofheinz2008}%
  \BibitemOpen
  \bibfield  {author} {\bibinfo {author} {\bibfnamefont {M.}~\bibnamefont
  {Hofheinz}}, \bibinfo {author} {\bibfnamefont {E.~M.}\ \bibnamefont {Weig}},
  \bibinfo {author} {\bibfnamefont {M.}~\bibnamefont {Ansmann}}, \bibinfo
  {author} {\bibfnamefont {R.~C.}\ \bibnamefont {Bialczak}}, \bibinfo {author}
  {\bibfnamefont {E.}~\bibnamefont {Lucero}}, \bibinfo {author} {\bibfnamefont
  {M.}~\bibnamefont {Neeley}}, \bibinfo {author} {\bibfnamefont {A.~D.}\
  \bibnamefont {O’Connell}}, \bibinfo {author} {\bibfnamefont
  {H.}~\bibnamefont {Wang}}, \bibinfo {author} {\bibfnamefont {J.~M.}\
  \bibnamefont {Martinis}},\ and\ \bibinfo {author} {\bibfnamefont {A.~N.}\
  \bibnamefont {Cleland}},\ }\bibfield  {title} {\bibinfo {title} {Generation
  of {F}ock states in a superconducting quantum circuit},\ }\href
  {https://doi.org/10.1038/nature07136} {\bibfield  {journal} {\bibinfo
  {journal} {Nature (London)}\ }\textbf {\bibinfo {volume} {454}},\ \bibinfo
  {pages} {310} (\bibinfo {year} {2008})}\BibitemShut {NoStop}%
\bibitem [{\citenamefont {Kurizki}\ \emph {et~al.}(2015)\citenamefont
  {Kurizki}, \citenamefont {Bertet}, \citenamefont {Kubo}, \citenamefont
  {Mølmer}, \citenamefont {Petrosyan}, \citenamefont {Rabl},\ and\
  \citenamefont {Schmiedmayer}}]{Kurizki2015}%
  \BibitemOpen
  \bibfield  {author} {\bibinfo {author} {\bibfnamefont {G.}~\bibnamefont
  {Kurizki}}, \bibinfo {author} {\bibfnamefont {P.}~\bibnamefont {Bertet}},
  \bibinfo {author} {\bibfnamefont {Y.}~\bibnamefont {Kubo}}, \bibinfo {author}
  {\bibfnamefont {K.}~\bibnamefont {Mølmer}}, \bibinfo {author} {\bibfnamefont
  {D.}~\bibnamefont {Petrosyan}}, \bibinfo {author} {\bibfnamefont
  {P.}~\bibnamefont {Rabl}},\ and\ \bibinfo {author} {\bibfnamefont
  {J.}~\bibnamefont {Schmiedmayer}},\ }\bibfield  {title} {\bibinfo {title}
  {Quantum technologies with hybrid systems},\ }\href
  {https://doi.org/10.1073/pnas.1419326112} {\bibfield  {journal} {\bibinfo
  {journal} {Proceedings of the National Academy of Sciences}\ }\textbf
  {\bibinfo {volume} {112}},\ \bibinfo {pages} {3866} (\bibinfo {year}
  {2015})}\BibitemShut {NoStop}%
\bibitem [{\citenamefont {Blais}\ \emph {et~al.}(2021)\citenamefont {Blais},
  \citenamefont {Grimsmo}, \citenamefont {Girvin},\ and\ \citenamefont
  {Wallraff}}]{Blais2021}%
  \BibitemOpen
  \bibfield  {author} {\bibinfo {author} {\bibfnamefont {A.}~\bibnamefont
  {Blais}}, \bibinfo {author} {\bibfnamefont {A.~L.}\ \bibnamefont {Grimsmo}},
  \bibinfo {author} {\bibfnamefont {S.~M.}\ \bibnamefont {Girvin}},\ and\
  \bibinfo {author} {\bibfnamefont {A.}~\bibnamefont {Wallraff}},\ }\bibfield
  {title} {\bibinfo {title} {Circuit quantum electrodynamics},\ }\href
  {https://doi.org/10.1103/RevModPhys.93.025005} {\bibfield  {journal}
  {\bibinfo  {journal} {Rev. Mod. Phys.}\ }\textbf {\bibinfo {volume} {93}},\
  \bibinfo {pages} {025005} (\bibinfo {year} {2021})}\BibitemShut {NoStop}%
\bibitem [{\citenamefont {Nakamura}\ \emph {et~al.}(2001)\citenamefont
  {Nakamura}, \citenamefont {Pashkin},\ and\ \citenamefont
  {Tsai}}]{Nakamura2001}%
  \BibitemOpen
  \bibfield  {author} {\bibinfo {author} {\bibfnamefont {Y.}~\bibnamefont
  {Nakamura}}, \bibinfo {author} {\bibfnamefont {Y.~A.}\ \bibnamefont
  {Pashkin}},\ and\ \bibinfo {author} {\bibfnamefont {J.~S.}\ \bibnamefont
  {Tsai}},\ }\bibfield  {title} {\bibinfo {title} {Rabi oscillations in a
  {J}osephson-junction charge two-level system},\ }\href
  {https://doi.org/10.1103/PhysRevLett.87.246601} {\bibfield  {journal}
  {\bibinfo  {journal} {Phys. Rev. Lett.}\ }\textbf {\bibinfo {volume} {87}},\
  \bibinfo {pages} {246601} (\bibinfo {year} {2001})}\BibitemShut {NoStop}%
\bibitem [{\citenamefont {Oliver}\ \emph {et~al.}(2005)\citenamefont {Oliver},
  \citenamefont {Yu}, \citenamefont {Lee}, \citenamefont {Berggren},
  \citenamefont {Levitov},\ and\ \citenamefont {Orlando}}]{Oliver2005}%
  \BibitemOpen
  \bibfield  {author} {\bibinfo {author} {\bibfnamefont {W.~D.}\ \bibnamefont
  {Oliver}}, \bibinfo {author} {\bibfnamefont {Y.}~\bibnamefont {Yu}}, \bibinfo
  {author} {\bibfnamefont {J.~C.}\ \bibnamefont {Lee}}, \bibinfo {author}
  {\bibfnamefont {K.~K.}\ \bibnamefont {Berggren}}, \bibinfo {author}
  {\bibfnamefont {L.~S.}\ \bibnamefont {Levitov}},\ and\ \bibinfo {author}
  {\bibfnamefont {T.~P.}\ \bibnamefont {Orlando}},\ }\bibfield  {title}
  {\bibinfo {title} {Mach-{Z}ehnder interferometry in a strongly driven
  superconducting qubit},\ }\href {https://doi.org/10.1126/SCIENCE.1119678}
  {\bibfield  {journal} {\bibinfo  {journal} {Science}\ }\textbf {\bibinfo
  {volume} {310}},\ \bibinfo {pages} {1653} (\bibinfo {year}
  {2005})}\BibitemShut {NoStop}%
\bibitem [{\citenamefont {Wilson}\ \emph {et~al.}(2007)\citenamefont {Wilson},
  \citenamefont {Duty}, \citenamefont {Persson}, \citenamefont {Sandberg},
  \citenamefont {Johansson},\ and\ \citenamefont {Delsing}}]{Wilson2007}%
  \BibitemOpen
  \bibfield  {author} {\bibinfo {author} {\bibfnamefont {C.}~\bibnamefont
  {Wilson}}, \bibinfo {author} {\bibfnamefont {T.}~\bibnamefont {Duty}},
  \bibinfo {author} {\bibfnamefont {F.}~\bibnamefont {Persson}}, \bibinfo
  {author} {\bibfnamefont {M.}~\bibnamefont {Sandberg}}, \bibinfo {author}
  {\bibfnamefont {G.}~\bibnamefont {Johansson}},\ and\ \bibinfo {author}
  {\bibfnamefont {P.}~\bibnamefont {Delsing}},\ }\bibfield  {title} {\bibinfo
  {title} {Coherence times of dressed states of a superconducting qubit under
  extreme driving},\ }\href {https://doi.org/10.1103/PhysRevLett.98.257003}
  {\bibfield  {journal} {\bibinfo  {journal} {Phys. Rev. Lett.}\ }\textbf
  {\bibinfo {volume} {98}},\ \bibinfo {pages} {257003} (\bibinfo {year}
  {2007})}\BibitemShut {NoStop}%
\bibitem [{\citenamefont {Pirkkalainen}\ \emph {et~al.}(2013)\citenamefont
  {Pirkkalainen}, \citenamefont {Cho}, \citenamefont {Li}, \citenamefont
  {Paraoanu}, \citenamefont {Hakonen},\ and\ \citenamefont
  {Sillanp{\"a}{\"a}}}]{Pirkkalainen2013}%
  \BibitemOpen
  \bibfield  {author} {\bibinfo {author} {\bibfnamefont {J.-M.}\ \bibnamefont
  {Pirkkalainen}}, \bibinfo {author} {\bibfnamefont {S.~U.}\ \bibnamefont
  {Cho}}, \bibinfo {author} {\bibfnamefont {J.}~\bibnamefont {Li}}, \bibinfo
  {author} {\bibfnamefont {G.~S.}\ \bibnamefont {Paraoanu}}, \bibinfo {author}
  {\bibfnamefont {P.~J.}\ \bibnamefont {Hakonen}},\ and\ \bibinfo {author}
  {\bibfnamefont {M.~A.}\ \bibnamefont {Sillanp{\"a}{\"a}}},\ }\bibfield
  {title} {\bibinfo {title} {Hybrid circuit cavity quantum electrodynamics with
  a micromechanical resonator},\ }\href {https://doi.org/10.1038/nature11821}
  {\bibfield  {journal} {\bibinfo  {journal} {Nature}\ }\textbf {\bibinfo
  {volume} {494}},\ \bibinfo {pages} {211} (\bibinfo {year}
  {2013})}\BibitemShut {NoStop}%
\bibitem [{\citenamefont {Niemczyk}\ \emph {et~al.}(2010)\citenamefont
  {Niemczyk}, \citenamefont {Deppe}, \citenamefont {Huebl}, \citenamefont
  {Menzel}, \citenamefont {Hocke}, \citenamefont {Schwarz}, \citenamefont
  {Garcia-Ripoll}, \citenamefont {Zueco}, \citenamefont {H\"{u}mer},
  \citenamefont {Solano}, \citenamefont {Marx},\ and\ \citenamefont
  {Gross}}]{Niemczyk2010}%
  \BibitemOpen
  \bibfield  {author} {\bibinfo {author} {\bibfnamefont {T.}~\bibnamefont
  {Niemczyk}}, \bibinfo {author} {\bibfnamefont {F.}~\bibnamefont {Deppe}},
  \bibinfo {author} {\bibfnamefont {H.}~\bibnamefont {Huebl}}, \bibinfo
  {author} {\bibfnamefont {E.~P.}\ \bibnamefont {Menzel}}, \bibinfo {author}
  {\bibfnamefont {F.}~\bibnamefont {Hocke}}, \bibinfo {author} {\bibfnamefont
  {M.~J.}\ \bibnamefont {Schwarz}}, \bibinfo {author} {\bibfnamefont {J.~J.}\
  \bibnamefont {Garcia-Ripoll}}, \bibinfo {author} {\bibfnamefont
  {D.}~\bibnamefont {Zueco}}, \bibinfo {author} {\bibfnamefont
  {T.}~\bibnamefont {H\"{u}mer}}, \bibinfo {author} {\bibfnamefont
  {E.}~\bibnamefont {Solano}}, \bibinfo {author} {\bibfnamefont
  {A.}~\bibnamefont {Marx}},\ and\ \bibinfo {author} {\bibfnamefont
  {R.}~\bibnamefont {Gross}},\ }\bibfield  {title} {\bibinfo {title} {Circuit
  quantum electrodynamics in the ultrastrong-coupling regime},\ }\href
  {https://doi.org/10.1038/nphys1730} {\bibfield  {journal} {\bibinfo
  {journal} {Nat. Phys.}\ }\textbf {\bibinfo {volume} {6}},\ \bibinfo {pages}
  {772} (\bibinfo {year} {2010})}\BibitemShut {NoStop}%
\bibitem [{\citenamefont {Forn-D\'{\i}az}\ \emph {et~al.}(2010)\citenamefont
  {Forn-D\'{\i}az}, \citenamefont {Lisenfeld}, \citenamefont {Marcos},
  \citenamefont {Garc\'{\i}a-Ripoll}, \citenamefont {Solano}, \citenamefont
  {Harmans},\ and\ \citenamefont {Mooij}}]{FornDiaz2010}%
  \BibitemOpen
  \bibfield  {author} {\bibinfo {author} {\bibfnamefont {P.}~\bibnamefont
  {Forn-D\'{\i}az}}, \bibinfo {author} {\bibfnamefont {J.}~\bibnamefont
  {Lisenfeld}}, \bibinfo {author} {\bibfnamefont {D.}~\bibnamefont {Marcos}},
  \bibinfo {author} {\bibfnamefont {J.~J.}\ \bibnamefont {Garc\'{\i}a-Ripoll}},
  \bibinfo {author} {\bibfnamefont {E.}~\bibnamefont {Solano}}, \bibinfo
  {author} {\bibfnamefont {C.~J. P.~M.}\ \bibnamefont {Harmans}},\ and\
  \bibinfo {author} {\bibfnamefont {J.~E.}\ \bibnamefont {Mooij}},\ }\bibfield
  {title} {\bibinfo {title} {Observation of the {B}loch-{S}iegert shift in a
  qubit-oscillator system in the ultrastrong coupling regime},\ }\href
  {https://doi.org/10.1103/PhysRevLett.105.237001} {\bibfield  {journal}
  {\bibinfo  {journal} {Phys. Rev. Lett.}\ }\textbf {\bibinfo {volume} {105}},\
  \bibinfo {pages} {237001} (\bibinfo {year} {2010})}\BibitemShut {NoStop}%
\bibitem [{\citenamefont {Forn-D\'{\i}az}\ \emph {et~al.}(2019)\citenamefont
  {Forn-D\'{\i}az}, \citenamefont {Lamata}, \citenamefont {Rico}, \citenamefont
  {Kono},\ and\ \citenamefont {Solano}}]{FornDiaz2019}%
  \BibitemOpen
  \bibfield  {author} {\bibinfo {author} {\bibfnamefont {P.}~\bibnamefont
  {Forn-D\'{\i}az}}, \bibinfo {author} {\bibfnamefont {L.}~\bibnamefont
  {Lamata}}, \bibinfo {author} {\bibfnamefont {E.}~\bibnamefont {Rico}},
  \bibinfo {author} {\bibfnamefont {J.}~\bibnamefont {Kono}},\ and\ \bibinfo
  {author} {\bibfnamefont {E.}~\bibnamefont {Solano}},\ }\bibfield  {title}
  {\bibinfo {title} {Ultrastrong coupling regimes of light-matter
  interaction},\ }\href {https://doi.org/10.1103/RevModPhys.91.025005}
  {\bibfield  {journal} {\bibinfo  {journal} {Rev. Mod. Phys.}\ }\textbf
  {\bibinfo {volume} {91}},\ \bibinfo {pages} {025005} (\bibinfo {year}
  {2019})}\BibitemShut {NoStop}%
\bibitem [{Note2()}]{Note2}%
  \BibitemOpen
  \bibinfo {note} {The notation here follows the standard conventions of
  quantum optics. In circuit QED, the bare energy of the two-level system is
  frequently defined in the $\protect \hat {\sigma }_x$ basis; this differs
  from the usual quantum optics form by a rotation about $\protect \hat {\sigma
  }_y$.}\BibitemShut {Stop}%
\bibitem [{sup()}]{supp}%
  \BibitemOpen
  \href@noop {} {}\bibinfo {howpublished}
  {\url{URL_will_be_inserted_by_publisher}}\BibitemShut {NoStop}%
\bibitem [{\citenamefont {Casta\~{n}os}(2019)}]{Castanos2019}%
  \BibitemOpen
  \bibfield  {author} {\bibinfo {author} {\bibfnamefont {L.~O.}\ \bibnamefont
  {Casta\~{n}os}},\ }\bibfield  {title} {\bibinfo {title} {Simple, analytic
  solutions of the semiclassical {R}abi model},\ }\href
  {https://doi.org/10.1016/j.optcom.2018.08.046} {\bibfield  {journal}
  {\bibinfo  {journal} {Opt. Comm.}\ }\textbf {\bibinfo {volume} {430}},\
  \bibinfo {pages} {176} (\bibinfo {year} {2019})}\BibitemShut {NoStop}%
\bibitem [{\citenamefont {Hausinger}\ and\ \citenamefont
  {Grifoni}(2010)}]{Hausinger2010b}%
  \BibitemOpen
  \bibfield  {author} {\bibinfo {author} {\bibfnamefont {J.}~\bibnamefont
  {Hausinger}}\ and\ \bibinfo {author} {\bibfnamefont {M.}~\bibnamefont
  {Grifoni}},\ }\bibfield  {title} {\bibinfo {title} {Dissipative two-level
  system under strong ac driving: A combination of {F}loquet and {V}an {V}leck
  perturbation theory},\ }\href {https://doi.org/10.1103/PhysRevA.81.022117}
  {\bibfield  {journal} {\bibinfo  {journal} {Phys. Rev. A}\ }\textbf {\bibinfo
  {volume} {81}},\ \bibinfo {pages} {022117} (\bibinfo {year}
  {2010})}\BibitemShut {NoStop}%
\bibitem [{Note3()}]{Note3}%
  \BibitemOpen
  \bibinfo {note} {The validity of the Rabi model as an approximation to the
  underlying light-matter interaction, on the grounds of gauge invariance, has
  been the subject of recent debate. This question is, however, beyond the
  scope of the present work; we simply take the quantum Rabi Hamiltonian as a
  given.}\BibitemShut {Stop}%
\bibitem [{Note4()}]{Note4}%
  \BibitemOpen
  \bibinfo {note} {The corresponding matrix elements within the interaction
  picture with respect to the field are given in Eq.~(S.36) of the
  Supplementary Material~\cite {supp}.}\BibitemShut {Stop}%
\bibitem [{Note5()}]{Note5}%
  \BibitemOpen
  \bibinfo {note} {For the time-independent Hamiltonian $\protect \hat {H}_q$
  in the lab frame, this approximation may be obtained as the lowest order in
  degenerate perturbation theory. Alternatively, in the interaction picture
  with respect to the field (Eq.~(S.36) of the Supplementary Material~\cite
  {supp}), the approximation consists of neglecting the time-dependent terms,
  which is directly equivalent to the lowest-order semiclassical approximation
  mentioned in the previous paragraph.}\BibitemShut {Stop}%
\bibitem [{\citenamefont {Shankar}(1994)}]{Shankar}%
  \BibitemOpen
  \bibfield  {author} {\bibinfo {author} {\bibfnamefont {R.}~\bibnamefont
  {Shankar}},\ }\href@noop {} {\emph {\bibinfo {title} {Principles of Quantum
  Mechanics}}},\ \bibinfo {edition} {2nd}\ ed.\ (\bibinfo  {publisher} {Plenum
  Press},\ \bibinfo {address} {New York},\ \bibinfo {year} {1994})\BibitemShut
  {NoStop}%
\bibitem [{\citenamefont {Brune}\ \emph {et~al.}(1996)\citenamefont {Brune},
  \citenamefont {Schmidt-Kaler}, \citenamefont {Maali}, \citenamefont {Dreyer},
  \citenamefont {Hagley}, \citenamefont {Raimond},\ and\ \citenamefont
  {Haroche}}]{Brune1996}%
  \BibitemOpen
  \bibfield  {author} {\bibinfo {author} {\bibfnamefont {M.}~\bibnamefont
  {Brune}}, \bibinfo {author} {\bibfnamefont {F.}~\bibnamefont
  {Schmidt-Kaler}}, \bibinfo {author} {\bibfnamefont {A.}~\bibnamefont
  {Maali}}, \bibinfo {author} {\bibfnamefont {J.}~\bibnamefont {Dreyer}},
  \bibinfo {author} {\bibfnamefont {E.}~\bibnamefont {Hagley}}, \bibinfo
  {author} {\bibfnamefont {J.~M.}\ \bibnamefont {Raimond}},\ and\ \bibinfo
  {author} {\bibfnamefont {S.}~\bibnamefont {Haroche}},\ }\bibfield  {title}
  {\bibinfo {title} {Quantum {R}abi oscillation: A direct test of field
  quantization in a cavity},\ }\href
  {https://doi.org/10.1103/PhysRevLett.76.1800} {\bibfield  {journal} {\bibinfo
   {journal} {Phys. Rev. Lett.}\ }\textbf {\bibinfo {volume} {76}},\ \bibinfo
  {pages} {1800} (\bibinfo {year} {1996})}\BibitemShut {NoStop}%
\bibitem [{\citenamefont {Gradstein}\ and\ \citenamefont
  {Ryzhik}(2015)}]{GradsteinRyzhik}%
  \BibitemOpen
  \bibfield  {author} {\bibinfo {author} {\bibfnamefont {E.~S.}\ \bibnamefont
  {Gradstein}}\ and\ \bibinfo {author} {\bibfnamefont {I.~M.}\ \bibnamefont
  {Ryzhik}},\ }\href {http://www.sciencedirect.com/science/book/9780123736376}
  {\emph {\bibinfo {title} {Table of integrals, sums, series, and products}}},\
  \bibinfo {edition} {8th}\ ed.,\ edited by\ \bibinfo {editor} {\bibfnamefont
  {D.}~\bibnamefont {Zwillinger}}\ (\bibinfo  {publisher} {Academic Press},\
  \bibinfo {address} {Waltham, MA},\ \bibinfo {year} {2015})\BibitemShut
  {NoStop}%
\bibitem [{\citenamefont {Armour}\ \emph {et~al.}(2013)\citenamefont {Armour},
  \citenamefont {Blencowe}, \citenamefont {Brahimi},\ and\ \citenamefont
  {Rimberg}}]{Armour2013}%
  \BibitemOpen
  \bibfield  {author} {\bibinfo {author} {\bibfnamefont {A.~D.}\ \bibnamefont
  {Armour}}, \bibinfo {author} {\bibfnamefont {M.~P.}\ \bibnamefont
  {Blencowe}}, \bibinfo {author} {\bibfnamefont {E.}~\bibnamefont {Brahimi}},\
  and\ \bibinfo {author} {\bibfnamefont {A.~J.}\ \bibnamefont {Rimberg}},\
  }\bibfield  {title} {\bibinfo {title} {Universal quantum fluctuations of a
  cavity mode driven by a {J}osephson junction},\ }\href
  {https://doi.org/10.1103/PhysRevLett.111.247001} {\bibfield  {journal}
  {\bibinfo  {journal} {Phys. Rev. Lett.}\ }\textbf {\bibinfo {volume} {111}},\
  \bibinfo {pages} {247001} (\bibinfo {year} {2013})}\BibitemShut {NoStop}%
\bibitem [{\citenamefont {Allen}\ and\ \citenamefont
  {Eberly}(1987)}]{AllenEberly}%
  \BibitemOpen
  \bibfield  {author} {\bibinfo {author} {\bibfnamefont {L.}~\bibnamefont
  {Allen}}\ and\ \bibinfo {author} {\bibfnamefont {J.~H.}\ \bibnamefont
  {Eberly}},\ }\href@noop {} {\emph {\bibinfo {title} {Optical Resonance and
  Two-Level Atoms}}}\ (\bibinfo  {publisher} {Dover Publications},\ \bibinfo
  {address} {New York},\ \bibinfo {year} {1987})\BibitemShut {NoStop}%
\bibitem [{\citenamefont {Klimov}\ and\ \citenamefont
  {Chumakov}(2009)}]{KlimovChumakov}%
  \BibitemOpen
  \bibfield  {author} {\bibinfo {author} {\bibfnamefont {A.~B.}\ \bibnamefont
  {Klimov}}\ and\ \bibinfo {author} {\bibfnamefont {S.~M.}\ \bibnamefont
  {Chumakov}},\ }\href {https://doi.org/10.1002/9783527624003} {\emph {\bibinfo
  {title} {A Group-Theoretical Approach to Quantum Optics: Models of Atom-Field
  Interactions}}}\ (\bibinfo  {publisher} {Wiley-VCH},\ \bibinfo {address}
  {Weinheim},\ \bibinfo {year} {2009})\BibitemShut {NoStop}%
\bibitem [{\citenamefont {Ashhab}(2017)}]{Ashhab2017}%
  \BibitemOpen
  \bibfield  {author} {\bibinfo {author} {\bibfnamefont {S.}~\bibnamefont
  {Ashhab}},\ }\bibfield  {title} {\bibinfo {title}
  {Landau-{Z}ener-{S}tueckelberg interferometry with driving fields in the
  quantum regime},\ }\href {https://doi.org/10.1088/1751-8121/aa5f6e}
  {\bibfield  {journal} {\bibinfo  {journal} {J. Phys. A}\ }\textbf {\bibinfo
  {volume} {50}},\ \bibinfo {pages} {134002} (\bibinfo {year}
  {2017})}\BibitemShut {NoStop}%
\bibitem [{Note6()}]{Note6}%
  \BibitemOpen
  \bibinfo {note} {The appearance of the exponential term further depends on
  the ordering of the field operators; see the Supplemental Material~\cite
  {supp} for an alternative derivation with a different operator ordering,
  which gives rise to a different semiclassical expression.}\BibitemShut
  {Stop}%
\bibitem [{\citenamefont {Mollow}(1975)}]{Mollow1975}%
  \BibitemOpen
  \bibfield  {author} {\bibinfo {author} {\bibfnamefont {B.~R.}\ \bibnamefont
  {Mollow}},\ }\bibfield  {title} {\bibinfo {title} {Pure-state analysis of
  resonant light scattering: Radiative damping, saturation, and multiphoton
  effects},\ }\href@noop {} {\bibfield  {journal} {\bibinfo  {journal} {Phys.
  Rev. A}\ }\textbf {\bibinfo {volume} {12}},\ \bibinfo {pages} {1919}
  (\bibinfo {year} {1975})}\BibitemShut {NoStop}%
\bibitem [{\citenamefont {Pegg}(1980)}]{Pegg1980}%
  \BibitemOpen
  \bibfield  {author} {\bibinfo {author} {\bibfnamefont {D.~T.}\ \bibnamefont
  {Pegg}},\ }\bibfield  {title} {\bibinfo {title} {Atomic spectroscopy:
  interaction of atoms with coherent fields},\ }in\ \href@noop {} {\emph
  {\bibinfo {booktitle} {Laser physics: proceedings of the Second New Zealand
  Summer School in Laser Physics}}},\ \bibinfo {address} {University of
  Waikato},\ \bibinfo {editor} {edited by\ \bibinfo {editor} {\bibfnamefont
  {D.~F.}\ \bibnamefont {Walls}}\ and\ \bibinfo {editor} {\bibfnamefont
  {J.~D.}\ \bibnamefont {Harvey}}}\ (\bibinfo  {publisher} {Academic Press},\
  \bibinfo {address} {Sydney},\ \bibinfo {year} {1980})\ pp.\ \bibinfo {pages}
  {33--61}\BibitemShut {NoStop}%
\bibitem [{\citenamefont {Knight}\ and\ \citenamefont
  {Radmore}(1982)}]{Knight1982}%
  \BibitemOpen
  \bibfield  {author} {\bibinfo {author} {\bibfnamefont {P.~L.}\ \bibnamefont
  {Knight}}\ and\ \bibinfo {author} {\bibfnamefont {P.~M.}\ \bibnamefont
  {Radmore}},\ }\bibfield  {title} {\bibinfo {title} {Quantum origin of
  dephasing and revivals in the coherent-state {J}aynes-{C}ummings model},\
  }\href@noop {} {\bibfield  {journal} {\bibinfo  {journal} {Phys. Rev. A}\
  }\textbf {\bibinfo {volume} {26}},\ \bibinfo {pages} {676} (\bibinfo {year}
  {1982})}\BibitemShut {NoStop}%
\bibitem [{\citenamefont {Berman}\ and\ \citenamefont
  {Ooi}(2014)}]{Berman2014}%
  \BibitemOpen
  \bibfield  {author} {\bibinfo {author} {\bibfnamefont {P.~R.}\ \bibnamefont
  {Berman}}\ and\ \bibinfo {author} {\bibfnamefont {C.~H.~R.}\ \bibnamefont
  {Ooi}},\ }\bibfield  {title} {\bibinfo {title} {Collapse and revivals in the
  {J}aynes-{C}ummings model: An analysis based on the {M}ollow
  transformation},\ }\href {https://doi.org/10.1103/PhysRevA.89.033845}
  {\bibfield  {journal} {\bibinfo  {journal} {Phys. Rev. A}\ }\textbf {\bibinfo
  {volume} {89}},\ \bibinfo {pages} {033845} (\bibinfo {year}
  {2014})}\BibitemShut {NoStop}%
\bibitem [{Note7()}]{Note7}%
  \BibitemOpen
  \bibinfo {note} {This result is implicit in the work of Knight and
  Radmore~\cite {Knight1982}, although the focus there is on the interplay
  between the classical and quantum interaction terms in the dynamics of the
  two-level system. Pegg~\cite {Pegg1980} mentions that neglecting the quantum
  term gives the semiclassical results, without consideration of the conditions
  under which such an approximation might be valid.}\BibitemShut {Stop}%
\bibitem [{Note8()}]{Note8}%
  \BibitemOpen
  \bibinfo {note} {The phase of $\alpha $ determines the phase of the
  sinusoidal classical drive; for example, choosing $\alpha $ real gives $\cos
  \omega _0 t$ as in Eq.~\protect \eqref {Hsc}}\BibitemShut {NoStop}%
\bibitem [{Note9()}]{Note9}%
  \BibitemOpen
  \bibinfo {note} {See Supplemental Material~\cite {[{See Supplemental Material
  at }][{ for derivations and a technical discussion of the limiting procedure
  in the Fock-state basis}]supp} for a further, more technical discussion of
  this point, based on the structure of the Hamiltonian.}\BibitemShut {Stop}%
\bibitem [{\citenamefont {Everitt}\ \emph {et~al.}(2009)\citenamefont
  {Everitt}, \citenamefont {Munro},\ and\ \citenamefont
  {Spiller}}]{Everitt2009}%
  \BibitemOpen
  \bibfield  {author} {\bibinfo {author} {\bibfnamefont {M.~J.}\ \bibnamefont
  {Everitt}}, \bibinfo {author} {\bibfnamefont {W.~J.}\ \bibnamefont {Munro}},\
  and\ \bibinfo {author} {\bibfnamefont {T.~P.}\ \bibnamefont {Spiller}},\
  }\bibfield  {title} {\bibinfo {title} {Quantum-classical crossover of a field
  mode},\ }\href {https://doi.org/10.1103/PhysRevA.79.032328} {\bibfield
  {journal} {\bibinfo  {journal} {Phys. Rev. A}\ }\textbf {\bibinfo {volume}
  {79}},\ \bibinfo {pages} {032328} (\bibinfo {year} {2009})}\BibitemShut
  {NoStop}%
\bibitem [{Note10()}]{Note10}%
  \BibitemOpen
  \bibinfo {note} {See Supplementary Material~\cite {[{See Supplemental
  Material at }][{ for derivations and a technical discussion of the limiting
  procedure in the Fock-state basis}]supp} for further discussion of this
  distinction and its importance.}\BibitemShut {Stop}%
\bibitem [{\citenamefont {Haroche}\ and\ \citenamefont
  {Raimond}(2006)}]{HarocheRaimond}%
  \BibitemOpen
  \bibfield  {author} {\bibinfo {author} {\bibfnamefont {S.}~\bibnamefont
  {Haroche}}\ and\ \bibinfo {author} {\bibfnamefont {J.-M.}\ \bibnamefont
  {Raimond}},\ }\href@noop {} {\emph {\bibinfo {title} {Exploring the Quantum:
  Atoms, Cavities, and Photons}}}\ (\bibinfo  {publisher} {Oxford University
  Press},\ \bibinfo {address} {Oxford},\ \bibinfo {year} {2006})\BibitemShut
  {NoStop}%
\bibitem [{\citenamefont {Hohenester}(2020)}]{Hohenester}%
  \BibitemOpen
  \bibfield  {author} {\bibinfo {author} {\bibfnamefont {U.}~\bibnamefont
  {Hohenester}},\ }\href@noop {} {\emph {\bibinfo {title} {Nano and Quantum
  Optics: An Introduction to Basic Principles and Theory}}}\ (\bibinfo
  {publisher} {Springer},\ \bibinfo {address} {Cham, Switzerland},\ \bibinfo
  {year} {2020})\BibitemShut {NoStop}%
\bibitem [{\citenamefont {Larson}\ and\ \citenamefont
  {Mavrogordatos}(2021)}]{Larson2021}%
  \BibitemOpen
  \bibfield  {author} {\bibinfo {author} {\bibfnamefont {J.}~\bibnamefont
  {Larson}}\ and\ \bibinfo {author} {\bibfnamefont {T.}~\bibnamefont
  {Mavrogordatos}},\ }\href {https://doi.org/10.1088/978-0-7503-3447-1} {\emph
  {\bibinfo {title} {The {J}aynes{\textendash}{C}ummings Model and Its
  Descendants}}},\ 2053-2563\ (\bibinfo  {publisher} {IOP Publishing},\
  \bibinfo {year} {2021})\BibitemShut {NoStop}%
\end{thebibliography}%


\begin{thebibliography}{7}%
\makeatletter
\providecommand \@ifxundefined [1]{%
 \@ifx{#1\undefined}
}%
\providecommand \@ifnum [1]{%
 \ifnum #1\expandafter \@firstoftwo
 \else \expandafter \@secondoftwo
 \fi
}%
\providecommand \@ifx [1]{%
 \ifx #1\expandafter \@firstoftwo
 \else \expandafter \@secondoftwo
 \fi
}%
\providecommand \natexlab [1]{#1}%
\providecommand \enquote  [1]{``#1''}%
\providecommand \bibnamefont  [1]{#1}%
\providecommand \bibfnamefont [1]{#1}%
\providecommand \citenamefont [1]{#1}%
\providecommand \href@noop [0]{\@secondoftwo}%
\providecommand \href [0]{\begingroup \@sanitize@url \@href}%
\providecommand \@href[1]{\@@startlink{#1}\@@href}%
\providecommand \@@href[1]{\endgroup#1\@@endlink}%
\providecommand \@sanitize@url [0]{\catcode `\\12\catcode `\$12\catcode
  `\&12\catcode `\#12\catcode `\^12\catcode `\_12\catcode `\%12\relax}%
\providecommand \@@startlink[1]{}%
\providecommand \@@endlink[0]{}%
\providecommand \url  [0]{\begingroup\@sanitize@url \@url }%
\providecommand \@url [1]{\endgroup\@href {#1}{\urlprefix }}%
\providecommand \urlprefix  [0]{URL }%
\providecommand \Eprint [0]{\href }%
\providecommand \doibase [0]{https://doi.org/}%
\providecommand \selectlanguage [0]{\@gobble}%
\providecommand \bibinfo  [0]{\@secondoftwo}%
\providecommand \bibfield  [0]{\@secondoftwo}%
\providecommand \translation [1]{[#1]}%
\providecommand \BibitemOpen [0]{}%
\providecommand \bibitemStop [0]{}%
\providecommand \bibitemNoStop [0]{.\EOS\space}%
\providecommand \EOS [0]{\spacefactor3000\relax}%
\providecommand \BibitemShut  [1]{\csname bibitem#1\endcsname}%
\let\auto@bib@innerbib\@empty
\bibitem [{\citenamefont {Mandel}\ and\ \citenamefont
  {Wolf}(1995)}]{MandelWolf}%
  \BibitemOpen
  \bibfield  {author} {\bibinfo {author} {\bibfnamefont {L.}~\bibnamefont
  {Mandel}}\ and\ \bibinfo {author} {\bibfnamefont {E.}~\bibnamefont {Wolf}},\
  }\href@noop {} {\emph {\bibinfo {title} {Optical Coherence and Quantum
  Optics}}}\ (\bibinfo  {publisher} {Cambridge University Press},\ \bibinfo
  {address} {Cambridge},\ \bibinfo {year} {1995})\BibitemShut {NoStop}%
\bibitem [{\citenamefont {Gradstein}\ and\ \citenamefont
  {Ryzhik}(2015)}]{GradsteinRyzhik}%
  \BibitemOpen
  \bibfield  {author} {\bibinfo {author} {\bibfnamefont {E.~S.}\ \bibnamefont
  {Gradstein}}\ and\ \bibinfo {author} {\bibfnamefont {I.~M.}\ \bibnamefont
  {Ryzhik}},\ }\href {http://www.sciencedirect.com/science/book/9780123736376}
  {\emph {\bibinfo {title} {Table of integrals, sums, series, and products}}},\
  \bibinfo {edition} {8th}\ ed.,\ edited by\ \bibinfo {editor} {\bibfnamefont
  {D.}~\bibnamefont {Zwillinger}}\ (\bibinfo  {publisher} {Academic Press},\
  \bibinfo {address} {Waltham, MA},\ \bibinfo {year} {2015})\BibitemShut
  {NoStop}%
\bibitem [{\citenamefont {Shirley}(1965)}]{Shirley1965}%
  \BibitemOpen
  \bibfield  {author} {\bibinfo {author} {\bibfnamefont {J.}~\bibnamefont
  {Shirley}},\ }\bibfield  {title} {\bibinfo {title} {Solution of the
  {S}chr\"{o}dinger equation with a {H}amiltonian periodic in time},\ }\href
  {https://doi.org/10.1103/PhysRev.138.B979} {\bibfield  {journal} {\bibinfo
  {journal} {Phys. Rev.}\ }\textbf {\bibinfo {volume} {138}},\ \bibinfo {pages}
  {B979} (\bibinfo {year} {1965})}\BibitemShut {NoStop}%
\bibitem [{\citenamefont {Polonsky}\ and\ \citenamefont
  {Cohen-Tannoudji}(1965)}]{Polonsky1965}%
  \BibitemOpen
  \bibfield  {author} {\bibinfo {author} {\bibfnamefont {N.}~\bibnamefont
  {Polonsky}}\ and\ \bibinfo {author} {\bibfnamefont {C.}~\bibnamefont
  {Cohen-Tannoudji}},\ }\bibfield  {title} {\bibinfo {title}
  {Interpr\'{e}tation quantique de la modulation de fr\'{e}quence},\ }\href
  {https://doi.org/10.1051/jphys:01965002607040900} {\bibfield  {journal}
  {\bibinfo  {journal} {J. Phys. (Paris)}\ }\textbf {\bibinfo {volume} {26}},\
  \bibinfo {pages} {409} (\bibinfo {year} {1965})}\BibitemShut {NoStop}%
\bibitem [{\citenamefont {Cohen-Tannoudji}\ \emph {et~al.}(2004)\citenamefont
  {Cohen-Tannoudji}, \citenamefont {Dupont-Roc},\ and\ \citenamefont
  {Grynberg}}]{AtomPhotonInteractions}%
  \BibitemOpen
  \bibfield  {author} {\bibinfo {author} {\bibfnamefont {C.}~\bibnamefont
  {Cohen-Tannoudji}}, \bibinfo {author} {\bibfnamefont {J.}~\bibnamefont
  {Dupont-Roc}},\ and\ \bibinfo {author} {\bibfnamefont {G.}~\bibnamefont
  {Grynberg}},\ }\href@noop {} {\emph {\bibinfo {title} {Atom-photon
  interactions: basic processes and applications}}}\ (\bibinfo  {publisher}
  {Wiley-VCH},\ \bibinfo {address} {Weinheim},\ \bibinfo {year}
  {2004})\BibitemShut {NoStop}%
\bibitem [{\citenamefont {Szeg\H{o}}(1975)}]{Szego}%
  \BibitemOpen
  \bibfield  {author} {\bibinfo {author} {\bibfnamefont {G.}~\bibnamefont
  {Szeg\H{o}}},\ }\href@noop {} {\emph {\bibinfo {title} {Orthogonal
  polynomials}}},\ \bibinfo {edition} {4th}\ ed.\ (\bibinfo  {publisher}
  {American Mathematical Society},\ \bibinfo {address} {Providence, Rhode
  Island},\ \bibinfo {year} {1975})\BibitemShut {NoStop}%
\bibitem [{\citenamefont {Ashhab}(2017)}]{Ashhab2017}%
  \BibitemOpen
  \bibfield  {author} {\bibinfo {author} {\bibfnamefont {S.}~\bibnamefont
  {Ashhab}},\ }\bibfield  {title} {\bibinfo {title}
  {Landau-{Z}ener-{S}tueckelberg interferometry with driving fields in the
  quantum regime},\ }\href {https://doi.org/10.1088/1751-8121/aa5f6e}
  {\bibfield  {journal} {\bibinfo  {journal} {J. Phys. A}\ }\textbf {\bibinfo
  {volume} {50}},\ \bibinfo {pages} {134002} (\bibinfo {year}
  {2017})}\BibitemShut {NoStop}%
\end{thebibliography}%

\end{document}


\title{Supplemental Material for \\
`Defining the semiclassical limit of the quantum Rabi Hamiltonian'}
\author{E. K. Twyeffort Irish}
\affiliation{School of Physics and Astronomy, University of Southampton, Highfield, Southampton, SO17 1BJ, United Kingdom}
\email{e.k.twyeffort@soton.ac.uk}
\author{A. D. Armour}
\affiliation{School of Physics and Astronomy and Centre for the Mathematics and Theoretical Physics of Quantum Non-Equilibrium Systems, University of Nottingham, Nottingham NG7 2RD, United Kingdom}
\date{\today}

\maketitle

\renewcommand{\thesection}{S.\Roman{section}}
\renewcommand{\theequation}{S.\arabic{equation}}

\section{Derivation of the semiclassical Bessel-function expansion}
Here we show how the Bessel-function form of the semiclassical Rabi Hamiltonian [Eq.~(3)] is derived using a unitary transformation. Transforming Eq.~(1) with the operator defined in Eq.~(2) produces
\begin{equation}
\begin{split}
\tilde{H}_{sc} &= \opdag{U_{sc}}(t) H_{sc} \op{U}_{sc}(t) - i \opdag{U_{sc}}(t) \frac{d \op{U}_{sc}(t)}{dt} \\
&= \tfrac{1}{2}\Omega \op{\sigma}_z \exp [-i(4A/\omega_0) \op{\sigma}_x \sin \omega_0 t] .
\end{split}
\end{equation}
Applying the Bessel function identity
\begin{equation}
\exp (iz \sin \theta) = \sum_{p = -\infty}^{\infty} J_p(z) e^{i p \theta} ,
\label{Bessel_expansion}
\end{equation}
$\tilde{H}_{sc}$ may then be written as
\begin{equation}
\tilde{H}_{sc} = \tfrac{1}{2} \Omega \op{\sigma}_z \sum_{p = -\infty}^{\infty} J_p\left[-\frac{4A}{\omega_0} \op{\sigma}_x\right] e^{i p \omega_0 t} .
\end{equation}
Since $J_p(z)$ contains only odd (even) powers of $z$ for $p$ odd (even), this simplifies to
\begin{equation}
\tilde{H}_{sc} = \tfrac{1}{2} \Omega \op{\sigma}_z \sum_{p = -\infty}^{\infty} J_{2p}(4A/\omega_0) e^{i(2p) \omega_0 t} - \tfrac{1}{2} \Omega \op{\sigma}_z \sum_{p = -\infty}^{\infty} J_{2p+1}(4A/\omega_0) e^{i(2p+1) \omega_0 t} \op{\sigma}_x .
\end{equation}
Finally, using the identity $J_{-p}(x) = (-1)^p J_p(x)$ for integer $p$, Eq.~(3) is obtained.

\section{Derivation of the quantum Bessel-function expansion}
To derive the Bessel-function expansion of the quantum Rabi model, we begin by transforming the Hamiltonian, Eq.~(4) with the spin-dependent displacement operator given in Eq.~(5):
\begin{equation}
\begin{split}
\tilde{H}_q &= \opdag{D}\left[-\frac{\lambda}{\omega_0} \op{\sigma}_x \right] H \op{D}\left[-\frac{\lambda}{\omega_0} \op{\sigma}_x \right] \\
&= \omega_0 \adag \a - \frac{\lambda^2}{\omega_0} + \tfrac{1}{2} \Omega \op{\sigma}_z \op{D} \left[-\frac{2 \lambda}{\omega_0} \op{\sigma}_x\right] .
\label{Hrabitrans}
\end{split}
\end{equation}
Using the Campbell-Baker-Hausdorff theorem, the remaining displacement operator in Eq.~\eqref{Hrabitrans} may be written in the normal-ordered form
\begin{equation}
\op{D} \left[-\frac{2 \lambda}{\omega_0} \op{\sigma}_x\right] = \exp \left[-\frac{2 \lambda}{\omega_0} \op{\sigma}_x (\adag - \a) \right]  = e^{-\chi^2/2} e^{-\chi \op{\sigma}_x \adag} e^{\chi \op{\sigma}_x \a} ,
\end{equation}
where $\chi \equiv 2 \lambda/\omega_0$. Expanding the exponentials, the transformed Hamiltonian becomes
\begin{equation}
\tilde{H}_q = \omega_0 \adag \a - \frac{\lambda^2}{\omega_0} + \tfrac{1}{2} \Omega \op{\sigma}_z e^{-\chi^2/2} \left[1 - \chi \op{\sigma}_x \adag + \frac{1}{2!} \chi^2 \adag^2 - \frac{1}{3!} \chi^3 \op{\sigma}_x \adag^3 + \dotso \right] \left[1 + \chi \op{\sigma}_x \a + \frac{1}{2!} \chi^2 \a^2 + \frac{1}{3!} \chi^3 \op{\sigma}_x \a^3 + \dotso \right] .
\end{equation}
The last term may be split into two sets of terms: one that depends on $\op{\sigma}_z$ and one that depends on $\op{\sigma}_y$. The Hamiltonian can then be written simply as
\begin{equation}
\tilde{H} = \omega_0 \adag \a - \frac{\omega_0 \chi^2}{4} + \tilde{H}_z + \tilde{H}_y ,
\label{splitHam}
\end{equation}
where
\begin{align}
\begin{split}
\tilde{H}_z &= \frac{1}{2} \Omega e^{-\chi^2/2} \op{\sigma}_z \biggl[1 - \chi^2 \adag \a + \frac{1}{2!} \chi^2 \adag^2 + \frac{1}{2!} \chi^2 \a^2 + \frac{1}{(2!)^2} \chi^4 \adag^2 \a^2 - \frac{1}{3!} \chi^4 \adag^3 \a - \frac{1}{3!} \chi^4 \adag \a^3 \\
& \qquad \qquad \qquad \quad + \frac{1}{4!} \chi^4 \adag^4 + \frac{1}{4!} \chi^4 \a^4 + \mathcal{O}(\chi^6) \biggr] , 
\end{split} \\
\begin{split}
\tilde{H}_y &= \frac{i}{2} \Omega e^{-\chi^2/2} \op{\sigma}_y \biggl[-\chi \adag + \chi \a + \frac{1}{2!} \chi^3 \adag^2 \a - \frac{1}{2!} \chi^3 \adag \a^2 - \frac{1}{3!} \chi^3 \adag^3 + \frac{1}{3!} \chi^3 \a^3 - \frac{1}{3!2!} \chi^5 \adag^3 \a^2 + \frac{1}{3!2!} \chi^5 \adag^2 \a^3  \\
& \qquad \qquad \qquad \qquad + \frac{1}{4!} \chi^5 \adag^4 \a - \frac{1}{4!} \chi^5 \adag \a^4 - \frac{1}{5!} \chi^5 \adag^5 + \frac{1}{5!} \chi^5 \a^5 + \mathcal{O}(\chi^7) \biggr] .
\end{split}
\end{align}
Let us now move to a rotating frame with respect to the field Hamiltonian, using the transformation $\op{U}(t) = \exp(i \omega_0 t \adag \a)$. In this frame, $\tilde{H}_z$ becomes
\begin{equation}
\begin{split}
\tilde{H}_z(t) &= \op{U}(t) \tilde{H}_z \opdag{U}(t) \\
&= \tfrac{1}{2} \Omega e^{-\chi^2/2} \op{\sigma}_z \biggl[1 - \chi^2 \adag \a + \frac{1}{2!} \chi^2 (\adag^2 e^{2i \omega_0 t} + \a^2 e^{-2i \omega_0 t}) + \frac{1}{(2!)^2} \chi^4 \adag^2 \a^2 - \frac{1}{3!} \chi^4 (\adag^3 \a e^{2i \omega_0 t} + \adag \a^3 e^{-2i \omega_0 t}) \\
& \qquad \qquad \qquad \quad + \frac{1}{4!} \chi^4 (\adag^4 e^{4i \omega_0 t} + \a^4 e^{-4i \omega_0 t}) + \mathcal{O}(\chi^6) \biggr] . 
\end{split}
\end{equation}
Collecting together terms with the same time dependence, it becomes evident that all terms with frequency $\pm n \omega_0$ create a net change of $n$ photons in the field. For example, the time-independent terms in the interaction picture correspond to zero-photon transitions (in the displaced basis), which simply modify the qubit frequency:
\begin{equation}
\begin{split}
\tilde{H}_z^{(0)}(t) &= \tfrac{1}{2} \Omega e^{-\chi^2/2} \op{\sigma}_z \biggl[1 - \chi^2 \adag \a + \frac{1}{(2!)^2} \chi^4 \adag^2 \a^2 + \frac{1}{(3!)^2} \chi^6 \adag^3 \a^3 + \dots \biggr] \\
&= \tfrac{1}{2} \Omega e^{-\chi^2/2} \op{\sigma}_z \sum_{m=0}^{\infty} \frac{(-1)^m \chi^{2m}}{(m!)^2} \adag^m \a^m .
\end{split}
\end{equation}
The sum can be evaluated in closed form using the identity 
\begin{equation}
J_p(z) = \sum_{m=0}^{\infty} \frac{(-1)^m (z/2)^{2m+p}}{m!(m+p)!} ,
\end{equation}
where $J_p(z)$ is a Bessel function of the first kind. The zero-photon Hamiltonian then becomes
\begin{equation}
\tilde{H}_z^{(0)}(t) = \tfrac{1}{2} \Omega e^{-\chi^2/2} \op{\sigma}_z \normord{J_0(2\chi \sqrt{\adag \a})} ,
\label{zerophoton}
\end{equation}
where $\normord{{}}$ denotes normal-ordering without the use of commutators, e.g. $\normord{\a \adag} = \adag \a$. In a similar fashion, the higher order terms in $\tilde{H}_z(t)$ can be written in terms of Bessel functions, giving
\begin{equation}
\tilde{H}_z(t) = \tfrac{1}{2} \Omega e^{-\chi^2/2} \op{\sigma}_z \normord{J_0(2 \chi \sqrt{\adag \a})} + \tfrac{1}{2} \Omega e^{-\chi^2/2} \op{\sigma}_z  \sum_{p=1}^{\infty} \normord{\frac{J_{2p}(2 \chi \sqrt{\adag \a})}{(\sqrt{\adag \a})^{2p}} [e^{2ip\omega_0 t} \adag^{2p} + e^{-2ip\omega_0 t}\a^{2p}]} .
\end{equation}
$\tilde{H}_y(t)$ may be treated in the same way, with the result
\begin{equation}
\tilde{H}_y(t) = -\tfrac{i}{2} \Omega e^{-\chi^2/2} \op{\sigma}_y  \sum_{p=0}^{\infty} \normord{\frac{J_{2p+1}(2 \chi \sqrt{\adag \a})}{(\sqrt{\adag \a})^{2p+1}} [e^{i(2p+1)\omega_0 t} \adag^{2p+1} - e^{-i(2p+1)\omega_0 t}\a^{2p+1}]} .
\end{equation}
These terms may now be put back into Eq.~\eqref{splitHam}. In the rotating frame, the $\omega_0 \adag \a$ term is eliminated from the Hamiltonian. Writing $i \op{\sigma}_y = \op{\sigma}_z \op{\sigma}_x$ and noting that $\op{\sigma}_x^p = 1$ for even $p$, the $\tilde{H}_z(t)$ and $\tilde{H}_y(t)$ terms may be combined into a single summation, giving the Hamiltonian
\begin{equation}
\tilde{H}_q(t) = -\frac{\omega_0 \chi^2}{4} + \tfrac{1}{2} \Omega e^{-\chi^2/2} \op{\sigma}_z \normord{J_0(2 \chi \sqrt{\adag \a})} + \tfrac{1}{2} \Omega e^{-\chi^2/2} \op{\sigma}_z  \sum_{p=1}^{\infty}(-\op{\sigma}_x)^p \normord{\frac{J_{p}(2 \chi \sqrt{\adag \a})}{(\sqrt{\adag \a})^{p}} [e^{ip\omega_0 t} \adag^{p} + (-1)^p e^{-ip\omega_0 t}\a^{p}]} .   
\end{equation}
Replacing $\chi$ by $2\lambda/\omega_0$ produces the final form given in Eq.~(7).

\section{Treating the displacement operator as a spin transformation}

The spin-dependent displacement operator given in Eq.~(5) may be interpreted equivalently as a field-dependent transformation on the spin operators. Writing it in the form
\begin{equation}
\begin{split}
\op{D} \left[-\frac{ \lambda}{\omega_0} \op{\sigma}_x\right] &= \cosh\left[-\frac{ \lambda}{\omega_0} \op{\sigma}_x(\adag - \a) \right] + \sinh\left[-\frac{ \lambda}{\omega_0} \op{\sigma}_x(\adag - \a) \right] \\
&= \cosh\left[-\frac{ \lambda}{\omega_0} (\adag - \a) \right] + \op{\sigma}_x \sinh\left[-\frac{ \lambda}{\omega_0} (\adag - \a) \right] ,
\end{split}
\end{equation}
the transformed Hamiltonian in the rotating frame may be expressed as
\begin{equation}
    \tilde{H}_q = -\frac{\lambda^2}{\omega_0} + \tfrac{1}{2} \Omega\op{\sigma}_z \cosh\left[-\frac{2 \lambda}{\omega_0} (e^{i \omega_0 t} \adag - e^{-i \omega_0 t} \a) \right] + \tfrac{i}{2} \Omega \op{\sigma}_y \sinh\left[-\frac{2 \lambda}{\omega_0} (e^{i \omega_0 t} \adag - e^{-i \omega_0 t} \a) \right] .
\end{equation}
Noting that the expectation value $\langle -(2\lambda/\omega_0) (e^{i \omega_0 t} \adag - e^{-i \omega_0 t} \a) \rangle$ is a purely imaginary number, this may be thought of as a rotation of the spin operators, where the angle of the rotation depends on the field. 

It is interesting to examine what happens when the `standard' procedure for deriving a corresponding semiclassical model is used on this form of the quantum Hamiltonian. Apply the replacement $\adag \to \alpha^*$, $\a \to \alpha$ and define $\alpha = \lvert \alpha \vert e^{-i \phi}$, so that 
\begin{equation}
    e^{i \omega_0 t} \adag - e^{-i \omega_0 t} \a \to 2i \lvert \alpha \rvert \sin \phi \sin(\omega_0 t) .
\end{equation}
Putting this back into the Hamiltonian and expanding the hyperbolic functions in terms of exponentials,
\begin{equation}
\begin{split}
    \tilde{H} &= -\frac{\lambda^2}{\omega_0} + \tfrac{1}{2} \Omega\op{\sigma}_z \left\{\exp[-4i \lambda \lvert \alpha \rvert \sin \phi \sin(\omega_0 t)] + \exp[4i \lambda \lvert \alpha \rvert \sin \phi \sin(\omega_0 t)] \right\} \\
    & \quad + \tfrac{i}{2} \Omega\op{\sigma}_y \left\{\exp[-4i \lambda \lvert \alpha \rvert \sin \phi \sin(\omega_0 t)] - \exp[4i \lambda \lvert \alpha \rvert \sin \phi \sin(\omega_0 t)] \right\} .
\end{split}
\end{equation}
Utilising the Bessel function identity \eqref{Bessel_expansion} and noting again that $J_p(-x) = (-1)^p J_p(x)$ and that $\op{\sigma}_z \op{\sigma}_x^p = \sigma_x$ for $p$ even and $i \op{\sigma}_y$ for $p$ odd, the Hamiltonian becomes
\begin{equation}
    \tilde{H} = -\frac{\lambda^2}{\omega_0} + \tfrac{1}{2} \Omega\op{\sigma}_z J_0(4 \lambda \lvert \alpha \rvert/\omega_0) + \tfrac{1}{2} \Omega\op{\sigma}_z \sum_{p = 1}^{\infty} (-\op{\sigma}_x)^p J_{p}(4 \lambda \lvert \alpha \rvert/\omega_0)[e^{i p \phi} e^{ip \omega_0 t} + (-1)^p e^{-i p \phi} e^{-ip \omega_0 t} ] .
\label{hyp_scBessel}
\end{equation}
Comparing with Eq.~(3), it is evident that $\phi = 0$ corresponds to the same choice of phase for the driving field as in the original semiclassical Hamiltonian, Eq.~(1). 

Interestingly, this form differs from both Eq.~(3) and Eq.~(7). The constant term $-\lambda^2/\omega_0$ originates from the application of the quantum displacement transformation and hence appears in Eqs.~\eqref{hyp_scBessel} and (7) but not in Eq.~(3), which was derived directly from the semiclassical Rabi Hamiltonian via the semiclassical transformation (2). As discussed in the main text, Eq.~(7) contains an additional factor of $e^{-2 \lambda^2/\omega_0^2}$ arising from the normal-ordering procedure. 

We see, then, that two mathematically equivalent forms of the quantum Hamiltonian, which differ only in the ordering of the quantum field operators, give rise to different semiclassical Hamiltonians when the standard recipe is followed. There is no {\em a priori} reason for preferring one ordering over another. One might suppose normal ordering to be associated with the semiclassical limit given that taking the expectation value $\bra{\alpha} f^{(N)}(\a, \adag) \ket{\alpha}$ of a normal-ordered function $f^{(N)}(\a, \adag)$ in a coherent state $\ket{\alpha}$ is equivalent to replacing $\a$ ($\adag$) by $\alpha$ ($\alpha^*$) \cite{MandelWolf}, but this is only indicative and does not provide a mathematically rigorous justification for using this replacement procedure to derive the semiclassical model. The discrepancies between the three forms may be resolved by letting $\lambda \to 0$. This further supports the assertion that a robust procedure for deriving a well-defined semiclassical limit of the quantum model must include taking the limit of vanishing quantum coupling.

\section{Derivation of matrix elements in the displaced Fock-state basis}
In this section we outline the derivation of the matrix elements of the transformed quantum Hamiltonian $\tilde{H}_q$ in the displaced Fock-state basis $\ket{\alpha, n}$. For simplicity, here we work with the time-independent version of the Hamiltonian.

Several identities and relations will be used in the following derivation. The displaced Fock state $\ket{\alpha, n}$ may be equivalently expressed as
\begin{equation}
\ket{\alpha, n} = \op{D}(\alpha) \ket{n} = \op{D}(\alpha) \frac{(\adag)^n}{\sqrt{n!}} \ket{0}.
\label{stateexpand}
\end{equation}
The power series form of the normal-ordered, operator-valued Bessel functions that appear in Eq.~(7) is
\begin{equation}
\normord{\frac{J_p(2 \chi \sqrt{\adag \a})}{(\sqrt{\adag \a})^p} \bigl[\adag^p + (-1)^p \a^p\bigr]} = \sum_{l=0}^{\infty} \frac{(-1)^l \chi^{2l+p}}{l!(l+p)!} \bigl[\adag^{l+p} \a^l + (-1)^p \adag^l \a^{l+p} \bigr] .
\label{normordBessel}
\end{equation}
A most useful identity for working with field operators in normal ordering is~\cite{MandelWolf}
\begin{equation}
\a^n f^{(N)}(\a, \adag) = \normord{\Bigl(\a + \frac{\partial}{\partial \adag}\Bigr)^n f^{(N)}(\a, \adag)} ,
\end{equation}
where $f^{(N)}(\a, \adag)$ denotes a normal-ordered function, i.e. all factors of $\adag$ appear to the left of all factors of $\a$. After some manipulation, the relation
\begin{equation}
\begin{split}
\a^m f^{(N)}(\a, \adag) \adag^n &= 
\normord{\Bigl(\a + \frac{\partial}{\partial \adag}\Bigr)^m \left[\Bigl(\adag + \frac{\partial}{\partial \a} \Bigr)^n f^{(N)}(\a, \adag) \right]} \\
&= \sum_{j=0}^m \sum_{k=0}^n \frac{m! n!}{j! (m-j)! k! (n-k)!} \normord{\frac{\partial^j}{\partial \adag^j} \Bigl(\adag^{n-k} \frac{\partial^k}{\partial \a^k} f^{(N)}(\a, \adag) \Bigr) \a^{m-j} } 
\label{normident}
\end{split}
\end{equation}
may be obtained. The expectation value of $f^{(N)}(\a, \adag)$ in a coherent state $\ket{\alpha}$ (including the vacuum state $\ket{0}$) may be computed by simply replacing $\a$ ($\adag$) by the coherent-state amplitude $\alpha$ ($\alpha^*)$:
\begin{equation}
\bra{\alpha} f^{(N)}(\a, \adag) \ket{\alpha} = f^{(N)}(\alpha, \alpha^*) .
\end{equation}
We wish to calculate the matrix element
\begin{equation}
T_p^{m,n} \equiv \bra{\alpha, m} \normord{\frac{J_p(2 \chi \sqrt{\adag \a})}{(\sqrt{\adag \a})^p} \bigl(\adag^p + (-1)^p \a^p\bigr)} \ket{\alpha, n} .
\end{equation}
Inserting the form of the state given in Eq.~\eqref{stateexpand} and the power series form of the operator from Eq.~\eqref{normordBessel}, then carrying out the displacement transformation on the inner set of operators, this becomes
\begin{equation}
\begin{split}
T_p^{m,n} &= \sum_{l=0}^{\infty} \frac{(-1)^l \chi^{2l+p}}{l! (l+p)! \sqrt{m! n!}} \bra{0} \a^m \bigl[ (\adag + \alpha^*)^{l+p} (\a + \alpha)^l + (-1)^p(\adag + \alpha^*)^l (\a + \alpha)^{l+p}  \bigr] \adag^n \ket{0} \\
&= T_1 + (-1)^p T_2 .
\end{split}
\label{Tpmn}
\end{equation}
From this point, the strategy is to put the operators into normal order and evaluate the expectation value in the state $\ket{0}$, then carry out the summations in order to express the matrix element in closed form. 

Beginning with $T_2$, applying Eq.~\eqref{normident} gives
\begin{equation}
\begin{split}
\bra{0} \a^m (\adag + \alpha^*)^l (\a + \alpha)^{l+p} \adag^n \ket{0} &= \bra{0} \sum_{j=0}^{m} \sum_{k=0}^{n} \frac{m! n!}{j! (m-j)! k! (n-k)!} \normord{\frac{\partial^j}{\partial \adag^j} \left[\adag^{n-k} \frac{\partial^k}{\partial \a^k} (\adag + \alpha^*)^l (\a + \alpha)^{l+p} \right] \a^{m-j}} \ket{0} \\
&= \bra{0} \sum_{k=0}^{n} \frac{n!}{k! (n-k)!} \frac{\partial^m}{\partial \adag^m} \left[\adag^{n-k}  (\adag + \alpha^*)^l \right] \frac{\partial^k}{\partial \a^k} (\a + \alpha)^{l+p} \ket{0} ,
\end{split}
\end{equation}
where the second line follows because the term $\a^{m-j}$ acting on $\ket{0}$ gives 0 unless $m=j$. Expanding the $\adag$ derivative term, 
\begin{multline}
\bra{0} \a^m (\adag + \alpha^*)^l (\a + \alpha)^{l+p} \adag^n \ket{0} \\
= \bra{0} \sum_{k=0}^{n} \frac{n!}{k! (n-k)!} \sum_{i=0}^m \frac{m!}{(m-i)!i!} \left[\frac{\partial^i}{\partial \adag^i} \adag^{n-k} \right] \left[\frac{\partial^{m-i}}{\partial \adag^{m-i}} (\adag + \alpha^*)^l \right] \left[ \frac{\partial^k}{\partial \a^k} (\a + \alpha)^{l+p} \right]\ket{0} .
\end{multline}
For $i > n-k$, the leftmost derivative evaluates to 0, whereas for $i < n-k$ there will be at least one $\adag$ acting on $\bra{0}$, causing it to vanish. Hence only the $i = n-k$ term contributes:
\begin{equation}
\bra{0} \a^m (\adag + \alpha^*)^l (\a + \alpha)^{l+p} \adag^n \ket{0} = \bra{0} \sum_{k=0}^{n} \frac{n! m!}{k! (n-k)! (m-n+k)!} \left[\frac{\partial^{m-n+k}}{\partial \adag^{m-n+k}} (\adag + \alpha^*)^l \right] \left[ \frac{\partial^k}{\partial \a^k} (\a + \alpha)^{l+p} \right]\ket{0} .
\end{equation}
Note that the derivative terms will equal zero if $m-n+k > l$ or $k > l+p$. This implies that the lower bound of the sum on $l$ will be equal to the smaller of $k+m-n$ and $k-p$ when the expectation value is inserted back into $T_2$. The derivatives may be readily evaluated and the expectation value calculated by replacing $\a$ and $\adag$ by 0. 

Taking the case $m-n > p$, we now have
\begin{equation}
T_2 = \sum_{l=k+m-n}^{\infty} \frac{(-1)^l \chi^{2l+p}}{l! (l+p)! \sqrt{m! n!}} \sum_{k=0}^{n} \frac{n! m!}{k! (n-k)! (m-n+k)!} \left[\frac{l!}{(l-m+n-k)!} (\alpha^*)^{l-m+n-k} \right] \left[\frac{(l+p)!}{(l+p-k)!} \alpha^{l+p-k} \right] .
\end{equation}  
Shifting indices to $j = l-k-m+n$ and rearranging factors leads to
\begin{equation}
\begin{split}
T_2 &= \sqrt{\frac{n!}{m!}} (-\chi)^{m-n} \frac{\alpha^{p+m-n}}{\abs{\alpha}^{p+m-n}} 
\sum_{k=0}^n \frac{(-1)^k m!}{k!(n-k)!(m-n+k)!} \chi^{2k}  \sum_{j=0}^{\infty} \frac{(-1)^j \chi^{2j+m-n+p}}{(j+p+m-n)! j!} (\alpha^* \alpha)^{j+(p+m-n)/2} \\
&= \sqrt{\frac{n!}{m!}} (-\chi)^{m-n} \frac{\alpha^{p+m-n}}{\abs{\alpha}^{p+m-n}} L_n^{m-n}(\chi^2) J_{p+m-n}(2\chi \abs{\alpha}) ,
\end{split}
\end{equation}
where the last line follows from the power series definitions of the Laguerre polynomials and Bessel functions of the first kind. For $p > m-n$, an identical expression is obtained as long as $m \geq n$. 

The calculation for $T_1$ is carried out in the same way, with the result
\begin{equation}
T_1 = \sqrt{\frac{n!}{m!}} \chi^{m-n} \frac{(\alpha^*)^{p-m+n}}{\abs{\alpha}^{p-m+n}} L_n^{m-n}(\chi^2) J_{p-m+n}(2\chi \abs{\alpha}) .
\end{equation}
Having worked out $T_p^{m,n} = T_1 + (-1)^p T_2$, the Hamiltonian matrix elements given in Eq.~(9) are easily written down.

\section{Examining the limiting procedure in the Fock-state basis}

To illustrate the importance of the choice of basis and the particular form of the limits, we examine whether a similar technique can be applied to the Hamiltonian in the basis of Fock states. In the rotating frame, Eq.~(6) becomes (omitting the constant term for simplicity)
\begin{equation}
\langle n+k \vert \op{D}^{\dag}\left(-\tfrac{\lambda}{\omega_0} \op{\sigma}_x \right)\op{H}_q(t) \op{D}\left(-\tfrac{\lambda}{\omega_0} \op{\sigma}_x \right) \vert n \rangle = \tfrac{1}{2} \Omega  \op{\sigma}_z \op{\sigma}_x^k e^{i k \omega_0 t} e^{-2 \lambda^2/\omega_0^2} \bigl(-\tfrac{2\lambda}{\omega_0}\bigr)^k \sqrt{\tfrac{n!}{(n+k)!}}  L_n^k\bigl(\tfrac{4\lambda^2}{\omega_0^2}\bigr) .
\label{fock_rot_frame}
\end{equation}
Scaling the coupling as $\lambda = A/\sqrt{n}$ and expanding the square root of the factorial as
\begin{equation}
\left[\frac{n!}{(n+k)!}\right]^{1/2} = n^{-k/2} \left[1-\frac{k(k+1)}{2n} + \mathcal{O}(n^{-2}) \right]^{1/2} ,
\end{equation}
we obtain
\begin{equation}
\langle n+k \vert \op{D}^{\dag}\left(-\tfrac{\lambda}{\omega_0} \op{\sigma}_x \right)\op{H}_q(t) \op{D}\left(-\tfrac{\lambda}{\omega_0} \op{\sigma}_x \right) \vert n \rangle = \tfrac{1}{2} \Omega  \op{\sigma}_z \op{\sigma}_x^k e^{i k \omega_0 t} e^{-2 A^2/n \omega_0^2} \left(-\tfrac{2A}{\sqrt{n} \omega_0}\right)^k n^{-k/2} \left[1 + O(n^{-1}) \right]^{1/2} L_n^k\bigl(\tfrac{4A^2}{n \omega_0^2}\bigr) .
\end{equation}
Using the asymptotic relation~\cite{GradsteinRyzhik} $\lim_{n \to \infty}[n^{-p} L_n^{p}(x/n)] = x^{-1/2 p} J_p(2 \sqrt{x})$, the limit of the matrix elements becomes
\begin{equation}
\lim_{n \to \infty} \langle n+k \vert \op{D}^{\dag}\left(-\tfrac{\lambda}{\omega_0} \op{\sigma}_x \right)\op{H}_q(t) \op{D}\left(-\tfrac{\lambda}{\omega_0} \op{\sigma}_x \right) \vert n \rangle = \tfrac{1}{2} \Omega  \op{\sigma}_z(- \op{\sigma}_x)^k e^{i k \omega_0 t} J_k\bigl(\tfrac{4A}{\omega_0}\bigr) .
\label{num_basis_matelems}
\end{equation}
While the resulting expressions take the same form as the terms in $\tilde{H}_{sc}(t)$, certain conceptual difficulties ensue. Unlike Eq.~(10), Eq.~\eqref{num_basis_matelems} is not  diagonal in the field states. Furthermore, the amplitude $A$ depends on the photon number as $\lambda \sqrt{n}$, rather than being a constant as one would expect for a semiclassical field. 

These points are typically addressed~\cite{Shirley1965, Polonsky1965, AtomPhotonInteractions} by assuming a field with a large average number of photons $\bar{n}$ and a strongly peaked distribution, such as a coherent state. For photon numbers $n \approx \bar{n}$, $\sqrt{n+k} = \sqrt{n} + k/2\sqrt{n} + \mathcal{O}(n^{-3/2}) \approx \sqrt{\bar{n}}$ for values of $k \ll \bar{n}$, so that $A$ becomes approximately independent of $n$. Furthermore, for a strongly peaked distribution, the amplitudes of the field components $\ket{n+k}$ drop off rapidly with $k$, so that in the limit of large $\bar{n}$ only the diagonal terms contribute substantially to the resulting dynamics of the two-level system. Polonsky and Cohen-Tannoudji~\cite{Polonsky1965} employed such reasoning to show that the semiclassical Bessel-function dynamics of a two-level system could be derived from the fully quantized model for an initial coherent state of the field, for large $\bar{n}$ and small $\lambda$. 

An additional complication in the Fock-state case is that the asymptotic relationship between the Laguerre polynomials and Bessel functions is not unique. Szeg\H{o}~\cite[Eq.~(8.22.4)]{Szego} gives the asymptotic formula
\begin{equation}
e^{-x/2} x^{p/2} L_n^p(x) = N^{-p/2} \frac{\Gamma(n + p + 1)}{n!} J_p[2(Nx)^{1/2}] + \mathcal{O}(n^{p/2-3/4}) ,
\end{equation}
where $N = n + (p+1)/2$. This produces a classical amplitude proportional to $\sqrt{n+(k+1)/2}$ rather than $\sqrt{n}$. Ashhab~\cite{Ashhab2017} deduced the same scaling based on numerical comparisons between the semiclassical and quantum Rabi frequencies for $k$-photon resonance, although he appears to have been unaware of Szeg\H{o}'s asymptotic expression. While it can certainly be argued that $\bar{n} \gg (1, k)$ in the semiclassical limit and hence the difference is negligible, the fact that the argument of the Bessel function depends not only on the (average) photon number but also on the photon number difference presents a further conceptual sticking point when working in the Fock-state basis.

In our approach, these problems are neatly avoided by working in the displaced basis and taking $\lvert \alpha \rvert$, rather than $n$ or $\bar{n}$, to infinity. All of the relevant statistical properties of the field are contained in the choice of basis, so no additional assumptions need to be imposed externally. As $\alpha$ is an arbitrary $c$-number parameter, there is no mathematical difficulty involved in taking its limit to infinity. The resulting Hamiltonian takes a separable form in which the quantum field component reduces to the identity; the formal structure of the Hilbert space is preserved, but the quantization of the field has no effect on the dynamics of the system.

\section{Deriving the semiclassical transformation from the quantum operator}
In the rotating frame, Eq.~(5) becomes
\begin{equation}
\tilde{D}\left(-\tfrac{\lambda}{\omega_0} \op{\sigma}_x \right) = \exp \left[-\tfrac{\lambda}{\omega_0} \op{\sigma}_x (e^{i \omega_0 t} \adag - e^{-i \omega_0 t}\a) \right] .
\label{disp_rot}
\end{equation}
Applying the displacement,
\begin{equation}
\op{D}^{\dag}(\alpha) \tilde{D}\left(-\tfrac{\lambda}{\omega_0} \op{\sigma}_x \right) \op{D}(\alpha) = \exp\left[-\tfrac{\lambda}{\omega_0} \op{\sigma}_x \left(\alpha^* e^{i \omega_0 t} - \alpha e^{-i \omega_0 t} \right) \right] \tilde{D}\left(-\tfrac{\lambda}{\omega_0} \op{\sigma}_x \right) .
\end{equation}
The matrix elements of $\tilde{D}\left(-\tfrac{\lambda}{\omega_0} \op{\sigma}_x \right)$ are next evaluated in the basis $\ket{n}\otimes\ket{\pm z}$, where $\ket{n}$ are Fock states of the field and $\op{\sigma}_x \ket{\pm x} = \pm \ket{\pm x}$:
\begin{align}
\bra{\pm x} \otimes \bra{n+k} \tilde{D}\left(-\tfrac{\lambda}{\omega_0} \op{\sigma}_x \right) \ket{n} \otimes \ket{\pm x} &= e^{\pm (\lambda/\omega_0)^2/2} \left(\pm \tfrac{\lambda}{\omega_0} \right)^{k} \sqrt{\tfrac{n!}{(n+k)!}} L_n^{k}\left(\tfrac{\lambda^2}{\omega_0^2}\right) , \label{disp_mat_elems}\\
\bra{\mp x} \otimes \bra{n+k} \tilde{D}\left(-\tfrac{\lambda}{\omega_0} \op{\sigma}_x \right) \ket{n} \otimes \ket{\pm x} &= 0 .
\end{align}
In the limit $\lambda \to 0$, the left-hand side of Eq.~\eqref{disp_mat_elems} reduces to $\delta_{k,0}$. Hence, the end result of following the procedure is 
\begin{equation}
\begin{split}
\lim_{\lambda \to 0, \lvert \alpha \rvert \to \infty} \op{D}^{\dag}(\alpha) \tilde{D}\left(-\tfrac{\lambda}{\omega_0} \op{\sigma}_x \right) \op{D}(\alpha) &= \exp\left[-\tfrac{\lambda}{\omega_0} \op{\sigma}_x \left(\alpha^* e^{i \omega_0 t} - \alpha e^{-i \omega_0 t} \right) \right] \otimes \sum_{n} \ket{n}\bra{n} \\
&= \op{U}_{sc}(t) \otimes \op{I}_f .
\end{split}
\end{equation}
The standard recipe for obtaining the semiclassical equations, applied to Eq.~\eqref{disp_rot}, also correctly yields the semiclassical transformation operator. However, it is interesting to note that if the quantum operator is first written in normal order, a naive application of the usual method leads to an additional factor $e^{-2 \lambda^2/\omega_0^2}$. Taking the limit $\lambda \to 0$ is then necessary to resolve the discrepancy (as tangentially noted in~\cite[pp. 487--8]{AtomPhotonInteractions}). No such ambiguity arises in the procedure outlined here.

\bibliography{Rabi_Bessel_paper}